\documentclass[iop, tighten, numberedappendix, letterpaper]{emulateapj}

\usepackage{verbatim}
\usepackage{subfigure}

\def\h2{$\rm H_2$}

\newcommand{\msun}{M$_{\odot}$}

\newcommand{\halpha}{H$\alpha$}
\newcommand{\hii}{H{\sc II}}
\newcommand{\hi}{H{\sc I}}

\newcommand{\hoi}{Ho~{\sc I}}
\newcommand{\hoii}{Ho~{\sc II}}
\newcommand{\hoix}{Ho~{\sc IX}}

\begin{document}

\shortauthors{Weisz et al.}
\title{The ACS Nearby Galaxy Survey Treasury {\sc VIII}.  The Global Star Formation Histories of 60 Dwarf Galaxies in the Local Volume\footnote{Based on observations made with the NASA/ESA Hubble Space Telescope, obtained from the Data Archive at the Space Telescope Science Institute, which is operated by the Association of Universities for Research in Astronomy, Inc., under NASA constract NAS 5-26555.}}

\author{Daniel R. Weisz\altaffilmark{1,2},
Julianne J. Dalcanton\altaffilmark{1},
Benjamin F. Williams\altaffilmark{1}, 
Karoline M. Gilbert\altaffilmark{1,9},
Evan D. Skillman\altaffilmark{2},
Anil C. Seth\altaffilmark{3},  
Andrew E. Dolphin\altaffilmark{4}, 
Kristen B.~W.~McQuinn\altaffilmark{2},
Stephanie M. Gogarten\altaffilmark{1}, 
Jon Holtzman\altaffilmark{5}, 
Keith Rosema\altaffilmark{1},
Andrew Cole\altaffilmark{6},
Igor D. Karachentsev\altaffilmark{7},
Dennis Zaritsky\altaffilmark{8}
}

\altaffiltext{1}{University of Washington; dweisz@astro.washington.edu}
\altaffiltext{2}{University of Minnesota}
\altaffiltext{3}{CfA Fellow, Harvard-Smithsonian Center for Astrophysics}
\altaffiltext{4}{Raytheon}
\altaffiltext{5}{New Mexico State University}
\altaffiltext{6}{University of Tasmania}
\altaffiltext{7}{Russian Academy of Sciences}
\altaffiltext{8}{Steward Observatory, University of Arizona}
\altaffiltext{9}{Hubble Fellow}

\begin{abstract}

We present uniformly measured star formation histories (SFHs) of 60 nearby ($D$ $\lesssim$ 4 Mpc) dwarf galaxies based on color-magnitude diagrams of resolved stellar populations from images taken with Hubble Space Telescope and analyzed as part of the ACS Nearby Galaxy Survey Treasury program (ANGST).  This volume-limited sample contains 12 dSph/dE, 5 dwarf spiral, 28 dIrr, 12 dSph/dIrr (transition), and 3 tidal dwarf galaxies.  The sample spans a range of $\sim$ 10 in $M_{B}$ and covers a wide range of environments, from highly interacting to truly isolated. From the best fit SFHs we find three significant results: (1) the average dwarf galaxy formed  $\gtrsim$ 50\% of its stars by z $\sim$ 2 and 60\% of its stars by z $\sim$ 1, regardless of  current morphological type; (2) the mean SFHs of dIs, dTrans, and dSphs are similar over most of cosmic time, and only begin to diverge a few Gyr ago, with the clearest differences between the three appearing during the most recent 1 Gyr; and (3) the SFHs are complex and the mean values are inconsistent with simple SFH models, e.g., single bursts, constant SFRs, or smooth, exponentially declining SFRs.  The mean SFHs are in general agreement with the cosmic SFH, although we observe offsets at intermediate times (z $\sim$ 1) that could be evidence that low mass systems experienced delayed star formation relative to more massive galaxies. The sample shows a strong density-morphology relationship, such that the dSphs in the sample are less isolated than dIs.  We find that the transition from a gas-rich to gas-poor galaxy cannot be solely due to internal mechanisms such as stellar feedback, and instead is likely the result of external mechanisms, e.g., ram pressure and tidal stripping and tidal forces.  The average transition dwarf galaxy is slightly less isolated and less gas-rich than the typical dwarf irregular.  Further, the transition dwarfs can be divided into two groups: interacting and gas-poor or isolated and gas-rich, suggesting two possible evolutionary pathways.

\end{abstract}


\section{Introduction}
\label{intro}

Dwarf galaxies have come to play an increasingly important role in understanding how galaxies form and evolve.  As the smallest, least luminous, and most common systems in the universe, dwarf galaxies span a wide range in physical characteristics and occupy a diverse set of environments \citep[e.g.,][]{zwi57, hod71, koo92, mar97, mat98, van00, kar04}, making them excellent laboratories for direct studies of cause and effect in galaxy evolution.  The low average masses and metallicities of dwarf galaxies suggest they may be the best available analogs to the seeds of hierarchical galaxy formation in the early universe.

A cohesive picture of the evolution of dwarf galaxies remains elusive.  Historically, evolutionary scenarios have often been considered in the context of a dual morphological classification, namely dwarf spheroidals (dSphs; we include dwarf ellipticals in this general category) and dwarf irregulars (dIs).  The former are classified based on a smooth morphology, with no observed knots of SF, and are generally found to be gas-poor.  The latter exhibit morphological evidence for current/recent SF activity and have a high gas fraction. A third, rarer class of so-called `transition' dwarf galaxies, have a high gas fraction yet little to no recent SF activity \citep[e.g.,][]{san91, ski95,mat98, mil01, dol05, you07, del08}.  These galaxies may be an evolutionary link between dIs and dSphs \citep[e.g.,][]{gre03} or simply could be ordinary dIs witnessed between massive star forming events \citep[e.g.,][]{ski03a}. Understanding the relationship between these three types of dwarf galaxies, specifically determining if and how dIs evolve into dSphs, is among the most pressing questions in dwarf galaxy evolution \citep[e.g.,][]{baa51, hod71, dek86, bin86, hod89, ski95, mat98, dek03, gre03, ric05, may06, orb08, kaz10}.

Resolved stellar populations have proven to be an incredibly powerful tool for observationally constraining scenarios of dwarf galaxy evolution. Past patterns of SF and chemical evolution are encoded in a galaxy's optical color-magnitude diagram (CMD).  To extract this information, a number of sophisticated algorithms \citep[e.g.,][]{tgf89, ts96, gabc96, m97, hol99, hvg99, har01,dol02, ia02, col07, yl07, apa09, cig10} have been developed to measure the star formation history (SFH), i.e., the star formation rate (SFR) as a function of time and metallicity, by comparing observed CMDs with those generated from models of stellar evolution.  The robustness of this method has become increasingly solidified with a variety of techniques and stellar models converging on consistent solutions for a range of galaxies \citep[e.g.,][]{ski02, ski03b, gza05, mon10, mon10b}.

Analysis of CMD-based SFHs have become particularly prevalent in studies of the formation and evolution of the Local Group (LG).  The LG contains $\sim$ 80 dwarf galaxies ranging from highly isolated to strongly interacting \citep[e.g.,][]{mat98, van00, tol09}.  Results from CMD-based SFHs of dwarf galaxies in the LG have revealed that both dSphs and dIs feature complex SFHs, typically with dominant stellar components older than 10 Gyr \citep[e.g.,][]{hod89, mat98, dol05, tol09}.  The SFHs of individual LG dwarf galaxies, as well as aggregate compilations, now serve as the basis for our understanding of the evolution of dwarf galaxies and have significantly advanced our knowledge of galactic group dynamics \citep[e.g.,][]{may01a, may01b, may06, orb08, tol09, may09, kaz10}. 

Beyond the LG, only a small subset of dwarf galaxies have explicitly measured CMD-based SFHs \citep[e.g.,][]{dol03,wei08, mcq09, mcq10}.  Ground-based observations of resolved stellar populations in more distant galaxies are challenging, due to the faintness of individual stars and the effects of photometric crowding.  However, observations from the Hubble Space Telescope (HST), particularly the Advanced Camera for Surveys \citep[ACS;][]{for98}, have revolutionized the field of resolved stellar populations, producing stunning CMDs of dwarf galaxies both in and beyond the LG. Outside the LG, past HST observations of resolved stellar populations have been fairly piecemeal, with individual or small sets of galaxies as the typical targets. Subsequent analysis often employed different methodologies or sought different science goals. This lack of uniformity also makes it challenging to place the results from LG studies in the context of the broader universe, as LG dwarf galaxies may not be representative of the larger dwarf galaxy population \citep[e.g.,][]{van00}.  

The ACS Nearby Galaxy Survey Treasury \citep[ANGST;][]{dal09} was designed to help remedy this situation.  Using a combination of new and archival imaging taken with the HST/ACS and Wide Field Planetary Camera 2 \citep[WFPC2;][]{hol95}, ANGST provides a uniformly reduced, multi-color photometric database of the resolved stellar populations of a volume-limited sample of nearby galaxies (D $\lesssim$ 4 Mpc) that are strictly outside the LG.  Of the $\sim$ 70 galaxies in this sample,  60 are dwarf galaxies that span a range of $\sim$ 10 in M$_{B}$ and that reside in both isolated field and strongly interacting group settings, providing an unbiased statistical sample in which to study the detailed properties dwarf galaxy formation and evolution.  

In this paper, we present the uniformly analyzed SFHs of 60 dwarf galaxies based on observations, photometry, and artificial star tests produced by the ANGST program.  The focus of this study is on the lifetime SFHs of the sample galaxies, with the recent ($<$ 1 Gyr) SFHs the subject of a separate paper \citep{wei10}. We first briefly review the sample selection, observations, and photometry in \S \ref{data}.  We then summarize the technique of measuring SFHs in \S \ref{sfhs}. In \S \ref{results}, we discuss and compare the resultant SFHs in the context of both dwarf galaxy formation and evolution. We then explore our results with respect to the morphology--density relationship in \S \ref{morph}.  Finally, we discuss the evolution of dwarf galaxies in the context of cosmology in \S \ref{cosmic}. Cosmological parameters used in this paper assume a standard WMAP-7 cosmology as detailed in \citet{jar10}.

\section{The Data}
\label{data}

In this section, we briefly summarize the selection of the ANGST dwarf galaxy sample along with the observations and photometry.  A more detailed discussion of the ANGST program can be found in \citet{dal09}.

\subsection{Selection and Final Sample}
\label{selection}

We constructed the initial list of ANGST galaxies based on the Catalog of Neighboring Galaxies  \citep{kar04}, consisting exclusively of galaxies located beyond the zero velocity surface of the LG \citep{van00}.  We selected potential targets with $|b|$ $>$ $20\,^{\circ}$ to avoid observational difficulties associated with low Galactic latitudes. By simulating CMDs and crowding limits, we found that a maximum distance of $\sim$ 3.5 Mpc provides the optimal balance between observational efficiency and achieving the program science goals.  Because the sample of galaxies within 3.5 Mpc contains predominantly field galaxies, we chose to extend the distance limits in the direction of the M81 Group. This extension added to the diversity of galaxies in the sample, while maintaining the goal of observational efficiency, due to the M81 Group's close proximity to the fiducial distance limit (D$_{M81}$ $\sim$ 3.6 Mpc) and low foreground extinction values. Similarly, we also included galaxies in the direction of the NGC 253 clump (D$_{N253}$ $\sim$ 3.9 Mpc) in the Sculptor Filament \citep{kar03}, further extending the range of environments probed by the sample, while maintaining strict volume limits for galaxy selection.

\begin{figure*}[]
\begin{center}
\epsscale{0.8}
\plotone{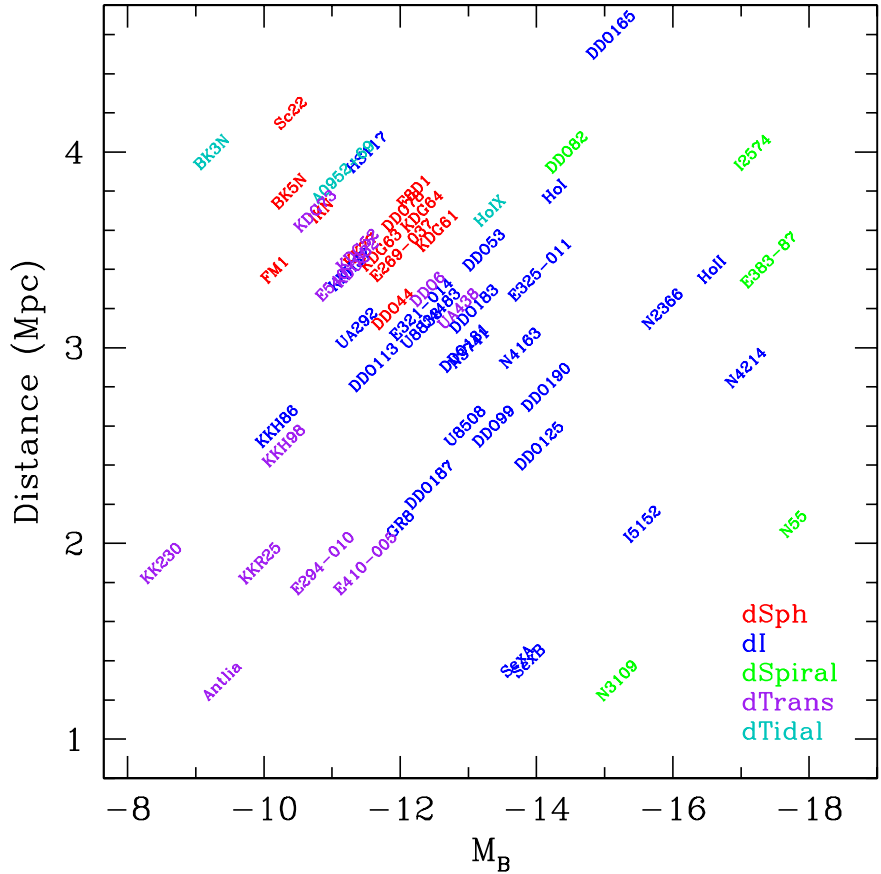}
\caption{Distribution of the ANGST sample of dwarf galaxies in distance and M$_{B}$.  Galaxies are color-coded by morphological type: dSphs (red; T $<$ 0), dIs (blue; T $=$ 10), dSpirals (green; T $=$ 8,9), dTrans (cyan; T $=$ 10), and dTidals (magenta; T $=$ 10) \citep{deV91, kar04}.  dTidals and dTrans have been identified based on previous literature analysis (see \S \ref{selection}).  We note there is likely some ambiguity in the classification of bright dIs and dSpirals (see \S \ref{selection}).}  
\label{sample}
\end{center}
\end{figure*}  

In this paper, we analyze a sample of 60 ANGST dwarf galaxies from both new and comparable quality archival imaging.  The galaxies range in M$_{B}$ from $-$8.23 (KK~230) to $-$17.77 (NGC~55) and in distance from 1.3 Mpc (Sex~A) to 4.6 Mpc (DDO~165).  We have included most galaxies considered dwarf galaxies in the literature in our analysis, although the upper mass/luminosity cutoff for what constitutes a dwarf galaxy is somewhat ambiguous \citep[e.g.,][]{hod71, ski96, mat98, tol09}.  For example, \citet{mat98} chose to exclude the LMC ($M_{B}=$ $-$17.93) and SMC ($M_{B}=$$-$16.35) from the `dwarfs' category.  Analogous to \citet{mat98}, we have not included NGC~3077 ($M_{B}=$$-$17.44), NGC~2976 ($M_{B}=$ $-$16.77), and NGC~300 ($M_{B}=$ $-$17.66), which would be among the brightest and most massive galaxies in the sample.  Detailed studies of the SFHs of NGC~2976 \citep{wil10} and NGC~300 \citep{gog10} are available in the literature.  A full list of sample galaxies and their properties are listed in Table \ref{tab1} with the distances and blue luminosities of the sample shown in Figure \ref{sample}.

Although the sample of ANGST dwarf galaxies is extensive, it is not complete within a fixed distance limit.  Galaxies at low galactic latitudes have been intentionally excluded from the original volume selection to avoid complications associated with high degrees of reddening.  While SFHs can still be derived from CMDs of such galaxies \citep[e.g., IC~4662;][]{mcq09,mcq10}, the effects of extreme reddening can lead to larger uncertainties and can require special analysis techniques (e.g., individual stellar line of sight reddening corrections), which detracts from a uniform approach to the data reduction.  In addition there have been a number of recently discovered dwarf galaxies in the M81 Group \citep{chi09}, which were not discovered in time to be included in the original ANGST sample.  The recent distance reassignment of UGC~4879 to the periphery of the LG \citep{kop08, jac10} meant this galaxy has also been excluded from the ANGST sample. Further, the ANGST sample is likely to be missing any faint dSphs which might be located in close proximity to M81, i.e., analogs to Milky Way satellites such as Draco and Ursa Minor, as such galaxies have not yet been detected due to their inherent faintness.  At a distance of M81 (3.6 Mpc) Draco and Ursa~Minor would have apparent blue luminosities of 18.98 and 20.13, respectively.  In comparison, Sc22 ($m_{B}$ $=$ 17.72) has the faintest apparent magnitude in the ANGST sample.

In this paper, we initially divide the sample of dwarf galaxies according to morphological type, $T$,  \citep{deV91, kar04} resulting in 12 dwarf galaxies with $T$ $\le$ 0 (dSphs), 5 with $T$ $=$ 8 or 9 (dwarf spirals; dSpirals), and 43 with $T$ $=$ 10 (dIs).  For ease of direct comparison with well studied LG dwarf galaxies, we have adopted nomenclature consistent with \citet{mat98}, in which dwarf ellipiticals (e.g., NGC~147, NGC~185) are found to be rare, and dSphs are more common.  Among the dIs, there could be some ambiguity between bright dIs and dSpirals.  For consistency, we defer to the $T$-type morphological classification scheme, but note that the distinction between a bright dI and dSpiral is not always clear. Morphological type $T$ $=$ 10 further includes transition dwarf galaxies (dTrans), galaxies with reduced recent star formation but high gas fractions \citep[e.g.,][]{mat98}, and tidal dwarf galaxy candidates (dTidals), which appear to be condensing out of tidally disturbed gas.   We classify subtypes, i.e., dTidal and dTrans, as follows: the three dTidals are Holmberg {\sc IX}, A0952+069, and BK3N, and are all located in the M81 Group.  For dTrans, we adopt the definition of \citet{mat98}, namely, that a galaxy has detectable gas but very little or no \halpha\ flux. The final sample of dTrans was classified based on \hi\ and \halpha\ measurements in the literature \citep{cot97, ski03a, kar04, kar07, beg08, bou09, cot09}.  We find 12 ANGST dwarf galaxies that satisfy the dTrans criteria: KK~230, Antlia, KKR~25, KKH~98, KDG~73, ESO294-10, ESO540-30, ESO540-32, KDG~52, ESO410-005, DDO~6, and UGCA~438, leaving the final tally of true dIs at 28.

\begin{deluxetable*}{lccccccccccc}
\tablecolumns{13}
\tabletypesize{\scriptsize}
\tablewidth{0pt}
\tablecaption{ANGST Dwarf Galaxy Sample}
\tablehead{
    \colhead{Galaxy} &
    \colhead{Alternate} &
    \colhead{Main} &
    \colhead{$M_{B}$} &
       \colhead{D} &
       \colhead{$A_{V}$} &
        \colhead{T} &
      \colhead{$\Theta$} &
        \colhead{$\mathcal{A}_{25}$} &
      \colhead{Filters} &
      \colhead{$M_{F814W}$ 50\%} &
    \colhead{HST} \\
        \colhead{Name} &
    \colhead{Name} &
    \colhead{Disturber} &
        \colhead{} &
       \colhead{(Mpc)} &
      \colhead{} &
      \colhead{} &
      \colhead{} &
      \colhead{} &
      \colhead{} &
         \colhead{Completeness} &
          \colhead{Proposal ID} \\
             \colhead{(1)} &
    \colhead{(2)} &
    \colhead{(3)} &
        \colhead{(4)} &
       \colhead{(5)} &
      \colhead{(6)} &
      \colhead{(7)} &
      \colhead{(8)} &
      \colhead{(9)} &
        \colhead{(10)} &
          \colhead{(11)} &
      \colhead{(12)} 
      }

\startdata

KK230 & KKR3 & M31 & -8.49 & 1.3 & 0.04 &  10 & -1.0 & 12.0 & F606W,F814W & $+$0.62 & 9771 \\
BK3N & & M81 &  -9.23 & 4.0 & 0.25 & 10 & 1.0 & 18.0 &  F475W,F814W & -0.32 & 10915 \\
Antlia & & N3109 & -9.38 & 1.3 & 0.24 & 10 & -0.1 & 1.20 &  F606W,F814W & $+$1.75 & 10210 \\
KKR25 & & M31 & -9.94 & 1.9 & 0.03 & 10 & -0.7 & 2.50 & F606W,F814W & -0.18 & 11986 \\
FM1 & F6D1 & M82 & -10.16 & 3.4 & 0.24 & -3 & 1.8 &  5.00 & F606W,F814W & $+$0.11 & 9884 \\
KKH86 & & M31 & -10.19 & 2.6 & 0.08 & 10 & -1.5 & 5.20 & F606W,F814W & -1.3 & 11986 \\
KKH98 &  & M31 &-10.29 & 2.5 & 0.39 & 10 & -0.7 & 5.40 & F475W,F814W & $+$0.46 & 10915 \\
BK5N--CENTRAL & & N3077 & -10.37 & 3.8  & 0.20 &-3 & 2.4 & 3.80 &  F606W,F814W & -0.77 & 6964 \\
BK5N--OUTER & & & &   & & &  & 3.80 &  F606W,F814W & -0.70 & 5898 \\
Sc22 & Sc-dE1 & N253 & -10.39 & 4.2 & 0.05 & -3 & 0.9 & 5.70 & F606W,F814W & $+$0.30 & 10503 \\
KDG73 & &M81 & -10.75 & 3.7 & 0.06 & 10 & 1.3 & 12.0 &  F475W,F814W & -0.38 &10915 \\
IKN & & M81 & -10.84 & 3.7 & 0.18 & -3 & 2.7 & 0.58 & F606W,F814W & -0.82 &  9771 \\
E294-010 & & N55 & -10.86 & 1.9 & 0.02 & -3 & 1.0 & 4.60 &  F606W,F814W & $+$1.80 & 10503 \\
A0952+69 & &N3077 & -11.16 & 3.9 & 0.26 &10 & 1.9 & 1.20 & F475W,F814W & -0.23 &10915 \\
E540-032 & & N253 &-11.22 & 3.4 & 0.06 & -3 & 0.6 & 2.30 & F606W,F814W & $+$0.10 & 10503 \\
KKH37 & & I342 & -11.26 & 3.4  & 0.23 &10 & -0.3 & 3.70 & F475W,F814W & $+$0.12 & 10915 \\
KDG2 & E540-030 & N253 & -11.29 & 3.4 & 0.07 & -1 & 0.4 & 2.70 & F606W,F814W & $+$0.32 &10503 \\
UA292 & CVnI-dwA &N4214 & -11.36 & 3.1 & 0.05 & 10 & -0.4 &  5.10 & F475W,F814W & -0.37 & 10915 \\
KDG52 & M81-Dwarf-A & M81 & -11.37 & 3.5 & 0.06 &10 & 0.7 & 2.30 & F555W,F814W & $+$0.37 &10605 \\
KK77 & F12D1 & M81 & -11.42 & 3.5 & 0.44 & -3 & 2.0 & 0.83 & F606W,F814W & $+$0.28 & 9884 \\
E410-005 & & N55 & -11.49 & 1.9 & 0.04 & -1 & 0.4 & 2.80 & F606W,F814W & $+$1.70 & 10503 \\
HS117 & & M81 & -11.51 & 4.0 & 0.36 & 10 & 1.9 & 2.70 & F606W,F814W & -0.81 & 9771 \\
DDO113 & UA276 & N4214 & -11.61 & 2.9 & 0.06 &10 & 1.6 & 1.80 &  F475W,F814W & $+$0.08 &10915 \\
KDG63 & U5428,DDO71 & M81 & -11.71 & 3.5 & 0.30 & -3 & 1.80 & 1.40 &  F606W,F814W & $+$0.31 & 9884 \\
DDO44 & UA133 & N2403 &  -11.89 & 3.2 & 0.13 & -3 & 1.7 & 0.59 & F475W,F814W & $+$0.16 & 10915 \\
GR8 & U8091,DDO155 &M31 & -12.00 & 2.1 & 0.08 & 10 & -1.2 & 3.30 & F475W,F814W & $+$0.82 & 10915 \\
E269-37 &  & N4945 & -12.02 & 3.5 & 0.44 & -3 & 1.6 & 3.80 & F606W,F814W & -1.8 & 11986 \\
DDO78 & & M81 & -12.04 & 3.7 &0.07 & -3 & 1.8 & 0.95 & F475W,F814W & -0.27 & 10915 \\
F8D1--CENTRAL & & M81 & -12.20 & 3.8 & 0.33 & -3 & 3.8 & 0.42 & F555W,F814W & -0.77 & 5898 \\
F8D1--OUTER & & & &  &  &  &  & 0.42 & F606W,F814W & -0.90 & 5898 \\
U8833 & & N4736 &-12.31 & 3.1 & 0.04 & 10 & -1.4 & 5.00 &  F606W,F814W & -0.23 & 10210 \\
E321-014 & & N5128 & -12.31 & 3.2 & 0.29 & 10 & -0.3 & 2.20 & F606W,F814W & -0.84 & 8601 \\
KDG64 & U5442 & M81 & -12.32 & 3.7 & 0.17 & -3 & 2.5 & 1.10 & F606W,F814W & $+$0.57 & 11986 \\
DDO6 & UA15 & N253 & -12.40 & 3.3 & 0.05 & 10 & 0.5 & 3.00 & F475W,F814W & -0.02 &10915 \\
DDO187 & U9128 & M31 & -12.43 & 2.3 & 0.07 & 10 & -1.3 & 1.60 &  F606W,F814W & $+$0.40 & 10210 \\
KDG61 & KK81 & M81 &-12.54 & 3.6 & 0.23 & -1 & 3.9 & 1.10 & F606W,F814W & $+$0.33 & 9884 \\
U4483 & & M81 & -12.58 & 3.2 & 0.11 & 10 & 0.5 & 2.20 & F555W,F814W & -1.36 & 8769 \\
UA438 & E407-18 & N55 & -12.85 & 2.2 & 0.05 & 10 & -0.7 & 1.00 & F606W,F814W & -1.7 & 8192 \\
DDO181 & U8651 & M81 &  -12.94 & 3.0 & 0.02 & 10 & -1.3 & 1.20 & F606W,F814W & -0.31 & 10210 \\
U8508 & IZw60 & M81 & -12.95 & 2.6 & 0.05 & 10 & -1.0 & 2.10 & F475W,F814W & $+$0.39 &10915 \\
N3741 & U6572 & M81 & -13.01 & 3.0 & 0.07 & 10 & -0.8 & 1.60 &  F475W,F814W & -0.24 & 10915 \\
DDO183 & U8760 & N4736 & -13.08 & 3.2 & 0.05 & 10 & -0.8 & 2.30 & F475W,F814W & -0.46 & 10915 \\
DDO53 & U4459 & M81 & -13.23 & 3.5 & 0.12 & 10 & 0.7 & 1.60 & F555W,F814W & -0.05 & 10605 \\
HoIX & U5336,DDO66 & M81 & -13.31 & 3.7 & 0.24 & 10 & 3.3 & 0.71 &  F555W,F814W & $+$0.13 &10605 \\
DDO99 & U6817 & N4214 & -13.37 & 2.6 & 0.08 & 10 & -0.5 & 0.58 & F606W,F814W & -0.95 & 10210 \\
SexA & DDO75 & MW & -13.71 & 1.3 & 0.14 & 10 & -0.6 & 0.06 & F555W,F814W & $+$0.80 & 7496 \\
N4163 & U7199 & N4190 & -13.76 & 3.0 & 0.06 & 10 & 0.1 & 1.20 & F475W,F814W & -0.04 & 10915 \\
SexB & U5373 & MW & -13.88 & 1.4 & 0.10 & 10 & -0.7 & 0.10 & F606W,F814W & $+$0.04 & 11986 \\
DDO125 & U7577 &  N4214 & -14.04 & 2.5 & 0.06 & 10 & -0.9 & 0.18 & F606W,F814W & -1.3 & 11986 \\
E325-11 &  & N5128 & -14.05 & 3.4 & 0.29 &10 & 1.1  & 0.52 & F606W,F814W & -0.58 &11986 \\
DDO190 & U9240 & M81 & -14.14 & 2.8 & 0.04 & 10 & -1.3 & 1.20 & F475W,F814W & -0.01 & 10915 \\
HoI & U5139,DDO63 & M81 &  -14.26 & 3.8 & 0.15 & 10 & 1.5 & 0.34 & F555W,F814W & $+$0.23 & 10605 \\
DDO82 & U5692 & M81 &  -14.44 & 4.0 & 0.13 & 9 & 0.9 & 0.53 & F475W,F814W & -0.32 & 10915 \\
DDO165 & U8201 & N4236 & -15.09 & 4.6 & 0.08  &10 & 0.0 & 0.50 & F555W,F814W & -0.7 & 10605 \\
N3109-DEEP & DDO236 & Antlia &  -15.18 & 1.3 & 0.20 & 9 & -0.1 & 0.05 & F606W,F814W & $+$0.50 & 10915 \\
N3109-WIDE2 & &  &  & & &  & & 0.05 & F606W,F814W & $+$0.09 & 11307 \\
I5152 &  & M31 & -15.55 & 2.1 & 0.08 & 10 & -1.1 & 0.10 &  F606W,F814W & -1.38 & 11986 \\
N2366--1 & U3851 & N2403 &  -15.85 & 3.2 & 0.11 & 10 & 1.0 & 0.39 & F555W,F814W & -0.16 & 10605 \\
N2366--2 & &  &  &  & &  &  & 0.39 & F555W,F814W & -0.05 & 10605 \\
\hoii--1& U4305 & M81 & -16.57 & 3.4 & 0.10 & 10 & 0.6 & 0.12 & F555W,F814W & -0.32 & 10605 \\
\hoii--2 & U4305 & &  &  &  &  &  & 0.12 & F555W,F814W & -0.21 & 10605 \\
N4214 & U7278 & DDO113 & -17.07 & 2.9 &0.07 & 10 & -0.7 & 0.03 &  F606W,F814W & -0.48 & 11986 \\
I2574--SGS & U5666,DDO81 & M81 &  -17.17 & 4.0 & 0.11 & 9 & 0.9 & 0.05 &  F555W,F814W & -0.26 & 9755 \\
I2574--1 & U5666,DDO81 &  & & &  &  &  & 0.05 &  F555W,F814W & -0.55 & 10605 \\
I2574--2 & U5666,DDO81 &  &  & &  &  &  & 0.05 &  F555W,F814W & -0.15 & 10605 \\
E383-87 &  & N5128 & -17.41 & 3.45 & 0.24 & 8 & -0.8 & 0.25 & F606W,F814W & -2.14 & 11986 \\
N55--CENTRAL & & N300 &  -17.77 & 2.1 & 0.04 & 8 & 0.4 & 0.01 &F606W,F814W & -1.55 & 9765 \\
N55--DISK & &  & &  & &  &  & 0.01 &F606W,F814W & -0.42 & 9765 
\enddata
\tablecomments{Properties of the sample of ANGST dwarf galaxies -- (1) \& (2) Names; (3) Most gravitationally influential neighbor \citep{kar04}; (4) Absolute Blue Magnitude \citep{kar04}; (5) TRGB Distance \citep{dal09, gil10}; (6) Foreground extinction \citep{sch98}; (7) Morphological T-Type \citep{deV91, kar04}; (8) Tidal Index \citep{kar04}; (9) Optical Coverage~Fraction: Angular area of ACS/WFPC2 coverage divided by angular area of the galaxy measured at a level of $\sim$ 25 mag arcsec$^{-2}$ or $\sim$ 26.5 mag arcsec$^{-2}$ for `KK' galaxies \citep{kar04}; (10) Filter Combination; (11)  50\% completeness limit ($M_{F814W}$); (12) HST Proposal ID, note new ANGST ACS (ID 10915) and WFC2 (11307, 11986) observations.}
\label{tab1}
\end{deluxetable*}

Throughout this paper, we adopt the tidal index, $\Theta$ \citep{kar04},  as a measure of a galaxy's isolation.  $\Theta$ describes the local mass density around galaxy $i$ as:

\begin{equation}
\Theta_{i} = max[log(M_{K}/D^{3}_{ik})] + C, \phantom{1}     i = 1,2,\cdots, N
\end{equation}

\noindent where $M_{k}$ is the total mass of any neighboring galaxy separated from galaxy $i$ by a distance of $D_{ik}$.  The values of $\Theta$ we use in this paper have been taken from \citet{kar04}.  Negative values correspond to more isolated galaxies, and positive values represent typical group members (see Table \ref{tab1}).

\subsection{Observations and Photometry}
\label{obs_phot}

HST observations of new ANGST targets were carried out in two phases due to the failure of ACS in 2007.  Prior to the failure, we observed new targets using ACS with WFPC2 in parallel mode.  Galaxies observed post-ACS failure were imaged with WFPC2 alone, as part of a `supplemental' HST program \citep[Proposal IDs 11307 and 11986 in Table \ref{tab1};][]{gil10}. New ACS observations used the F475W (SDSS $g^{\prime}$) and F814W ($I$) filter combination,  to optimize both photometric depth and temperature (color) baseline.  A third filter in F606W (wide $V$) is also available for most galaxies.  The low throughput in the bluer filters of WFPC2 led us to only use F606W and F814W for WFPC2 observations.

As described in \citet{dal09}, both new and archival observations were processed uniformly beginning with image reduction via the standard HST pipeline.  Using the ANGST data reduction pipeline, we performed photometry on each image with HSTPHOT\footnote[2]{http://purcell.as.arizona.edu/hstphot} \citep{dol00}, designed for WFPC2 images, and DOLPHOT\footnote[3]{http://purcell.as.arizona.edu/dolphot}, running in its ACS-optimized mode, for ACS observations, providing for a uniform treatment of all data.  The resultant photometry for each data set was filtered to ensure the final photometric catalogs excluded non-stellar objects such as cosmic rays, hot pixels, and extended sources.  For the purposes of this paper, we considered a star well measured if it met the following criteria: a signal-to-noise ratio $>$ 4 in both filters, a sharpness value such that ($sharp_{1}$ $+$ $sharp_{2}$)$^{2}$ $\le$ 0.075, and a crowding parameter such that ($crowd_{1}$ $+$ $crowd_{2}$) $\le$ 0.1.  To characterize observational uncertainties, we performed 500,000 artificial star tests on each image. Both the full and filtered (i.e., the `{\tt gst}' files) photometric catalogs, HST reference images, and CMDs are publicly available on MAST\footnote[4]{http://archive.stsci.edu/prepds/angst/}.  Definitions and detailed descriptions of the filtering criteria  and the observational strategies can be found in \citet{dol00}, \citet{dal09}, and \citet{gil10}.

Because of the variety of distances in the sample, the ACS/WFPC2 field of view does not subtend the same physical area for each galaxy. For certain comparisons among the sample (e.g., integrated stellar masses), it is important to account for these differences in coverage area.  To that effect,  we employ a simple areal normalization factor based on each galaxy's apparent blue surface brightness.  From the measurements listed in Table 4 of \citet{kar04}, we consider an effective elliptical area computed using the angular diameter and angular axial ratios at a blue surface brightness level of $\sim$ 25 mag arcsec$^{-2}$  ($\sim$ 26.5 mag arcsec$^{-2}$ in the case of the faintest galaxies).  We then calculate the normalization factor, $\mathcal{A}_{25}$, for each galaxy by taking the ratio of the angular area subtended by the ACS/ WFPC2 field of view to the angular area computed from \citet{kar04}.  This normalization has been specifically applied to the integrated stellar masses throughout this study.  The $\mathcal{A}_{25}$ normalizations are typically $>$ 1 (i.e., the HST field of view exceeds the area computed by \citealt{kar04}; see Table \ref{tab1}), and the most and least covered galaxies are BK3N (18.0) and NGC~55 (0.02), respectively. 

In most cases, a single HST field was sufficient to cover the main optical body of a galaxy, ensuring the SFHs are representative of the whole galaxy.  However, several of the sample galaxies required multiple observations to cover a reasonable fraction of the optical body (NGC~2366, Holmberg {\sc  II}, IC~2574, NGC~55, BK5N, F8D1, NGC~3109).  To derive the SFHs for each of these galaxies, we first checked to see that the 50\% completeness limits for each of the fields were similar (i.e., within $\pm$0.2 mag).  This condition was met for all galaxies except IC~2574, NGC~55, and NGC~3109, which have large angular sizes and wide ranges of surface brightnesses within each galaxy.  For galaxies with similar completeness limits, the photometry and false stars were first combined, and then the SFH code was run on the combined data.  For galaxies with overlapping fields (NGC~2366, Holmberg {\sc  II}) we carefully removed duplicate stars before merging the photometry.  IC~2574 was a special case as it has three overlapping fields with significantly different completeness limits.  In this instance, we first removed the duplicate stars from overlapping fields, ran the SFHs on each field, and then combined the results.  

NGC~55 and NGC~3109 have many new and archival HST observations taken over multiple visits.  In each case, we selected representative non-overlapping fields, one in the center of the galaxy and one in the disk.  The completeness functions were sufficiently different such that we did not combine photometry prior to computing the SFH.  Instead, we derived the SFHs for each field separately, and combined the resultant SFHs from each field.  The fractional coverage listed in Table \ref{tab1} takes into account the combined areas for galaxies with multiple fields.

\section{Method of Measuring SFHs}
\label{sfhs}

To ensure uniformity in the SFHs of the ANGST dwarf galaxies, we selected one SFH code \citep{dol02} and one set of stellar evolution models \citep[Padova;][]{mar08}.   The SFH code of \citet{dol02} provides the user with robust controls over critical fixed input variables, e.g., IMF, binary fraction, time and CMD bin sizes, as well as the ability to search for the combination of metallicity, distance, and extinction values that produce a model CMD that best fits the observed CMD.  The models of \citet{mar08} combine updated asymptotic giant branch (AGB) evolution tracks with the models of \citet{ber94} ($M$ $\ge$ 7 \msun) and \citet{gir02} ($M$ $<$ 7 \msun).  As with any models, there may be inevitable systematic biases associated with these particular choices \citep[e.g., at some metallicities the Padova model RGBs have red offsets from observations; e.g.,][]{gza05}.  However, these systematic effects will be shared by all galaxies in the sample, making for a robust relative comparison within the sample.  We discuss systematic uncertainties due to choice of isochrones in Appendix \ref{systematics}.  

Here, we briefly summarize the technique of measuring a SFH based on the full methodology described in \citet{dol02}.  The user specifies an assumed IMF and binary fraction, and allowable ranges in age, metallicity, distance, and extinction.  Photometric errors and completeness are characterized by artificial star tests.  From these inputs, many synthetic CMDs are generated to span the desired age and metallicity range.  For this work, we have used synthetic CMDs sampling stars with age and metallicity spreads of 0.1 and 0.1 dex, respectively.  These individual synthetic CMDs are then linearly combined along with a model foreground CMD to produce a composite synthetic CMD.  The linear weights on the individual CMDs are adjusted to obtain the best fit as measured by a Poisson maximum likelihood statistic; the weights corresponding to the best fit are the most probable SFH.  This process can be repeated at a variety of distance and extinction values to solve for these parameters as well.

Monte Carlo tests were used to estimate uncertainties due to both random and systematic sources.  For each Monte Carlo run, a Poisson random noise generator was used to create a random sampling of the best-fit CMD.  This CMD was then processed identically to the original solution, with additive errors in $M_{bol}$ and $\log(T_{eff})$ introduced when generating the model CMDs for these solutions.  Single shifts in $M_{bol}$ and $\log(T_{eff})$) are used for each Monte Carlo draw, and the errors themselves were drawn from normal distributions with 1-$\sigma$ values of 0.41 ($M_{bol}$) and 0.03 ($\log(T_{eff})$).  These distributions were designed to mimic the scatter in SFH uncertainties obtained by using multiple isochrone sets to fit the data, though we prefer to use this technique due to the small number of isochrone sets that fully cover the range of ages, metallicities, and evolutionary states required to adequately model our CMDs.  We found that the uncertainties were stable after 50 Monte Carlo tests, and thus conducted 50 realizations for each galaxy.  This technique of estimating error on SFHs is described in greater detail in \citet{dol10}. 

To minimize systematic effects for comparisons among the sample, we selected consistent parameters for measuring SFHs of all galaxies.  All SFHs were measured using a single slope power law IMF with a spectral index of $-$1.30 over a mass range of 0.1 to 120 \msun, a binary fraction of 0.35 with a flat secondary mass distribution, 71 equally spaced logarithmic time bins ranging from $\log{t}$ $=$ 6.6--10.15, color and magnitude bins of 0.05 and 0.1 mag, and the Padova stellar evolution models \citep{mar08}.  We note that the difference between our selected IMF and a Kroupa IMF \citep{kro01} is negligible, as the ANGST CMDs are limited to stellar masses $\gtrsim$ 0.8 \msun.  We designated the faint photometric limit to be equal to the 50\% completeness limit in each filter (see Table \ref{tab1}) as determined by $\sim$ 500,000 artificial star tests run for each galaxy.  The SFH program was initially allowed to search for the best fit distance and extinction values without constraints. Initial values for the distances were taken from the TRGB distances measured in \citet{dal09}, while foreground extinction values were taken from the galactic maps of \citet{sch98}.  We found no significant discrepancies between the assumed values and the best CMD fit distance and foreground extinction values, i.e., all were consistent within error.  Final solutions were computed using the TRGB distances from \citet{dal09} and \citet{gil10} and foreground extinction values from \citet{sch98}.

\begin{figure*}[]
\begin{center}
\epsscale{0.8}
\plotone{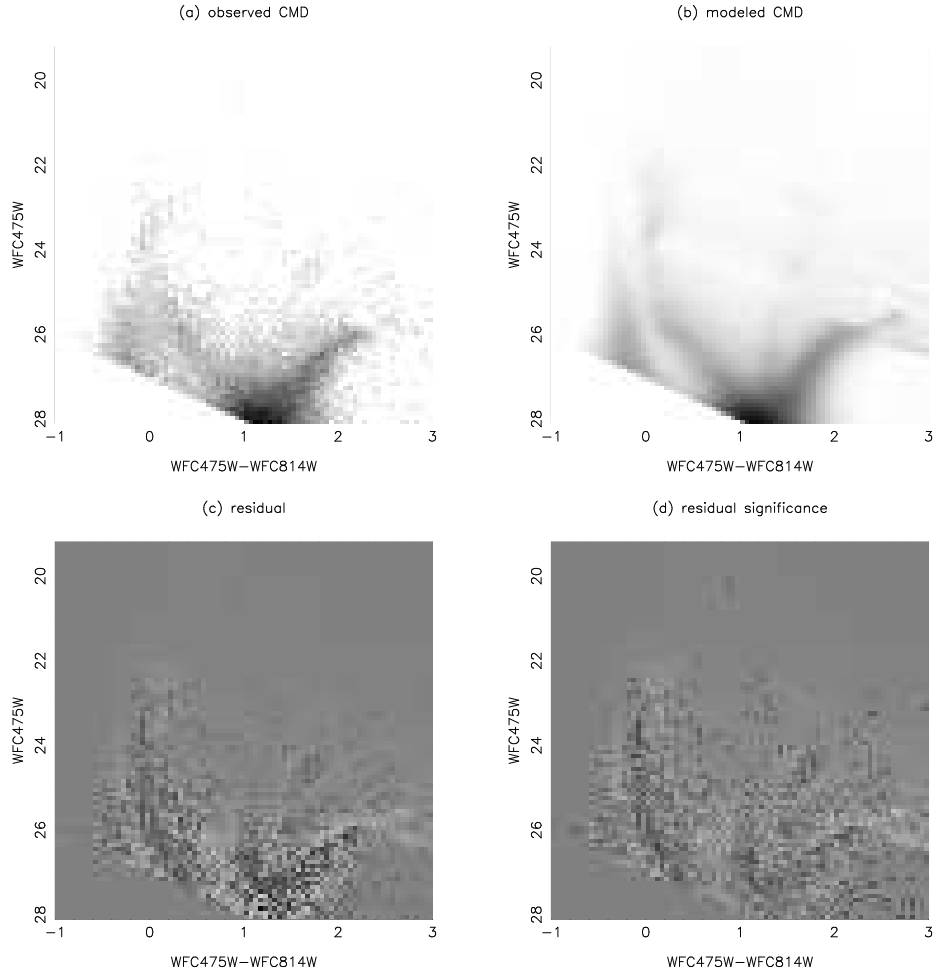}
\caption{A comparison of the observed and model CMDs for ANGST sample dI DDO~6.  The observed CMD is shown in panel (a) and the best matched model CMD in panel (b).  The lower two panels are diagnostic CMDs used to determine the fit quality.  Panel (c) shows the residual of data $-$ model CMD, with black and white points representing $\pm$ 5$\sigma$.  Panel (d) is the residual significance CMD, which is the data$-$model weighted by the variance in each CMD bin \citep{dol02}. Based on the residual significance CMD, we see that the overall fit is quite good, with only minor discrepancies, most notably between the main sequence and the blue helium burning stars, which is likely due to differential extinction effects, but only affects the SFH on time scales $<$ 1 Gyr.}  
\label{fit}
\end{center}
\end{figure*}

We placed an additional constraint on the CMD fitting process, namely that the mean metallicity in each time bin must monotonically increase toward the present.  The deepest ANGST CMDs do not reach the ancient MS turnoff, a requisite feature for completely breaking the age-metallicity degeneracy of the RGB only using broadband photometry \citep[e.g.,][]{col05, gza05}.   As a result, SFHs derived from shallow CMDs without a metallicity constraint can often have accompanying chemical evolution models that are unphysical (e.g., a drop in metallicity of several tenths of a dex over sub-Gyr time scales).  A more robust analysis of the chemical enrichment of dwarf galaxies would need to include the ancient MS, measured gas phase abundances, and/or individual stellar spectra \citep[e.g.,][]{col07, mon10, mon10b, kir10}.  We therefore consider analysis of the metallicity evolution of the ANGST sample beyond the scope of this paper.

As an example of a typical CMD fit made by the SFH code, we compare the observed and synthetic CMDs of a representative ANGST dTrans, DDO~6 (Figure \ref{fit}).  \citet{dal09} determined the TRGB distance of DDO~6 to be 3.31$\pm$0.06 Mpc while the foreground extinction maps of \citet{sch98} give values of A$_{B}$ $=$ 0.07 and A$_{V}$ $=$ 0.06.  Allowing the SFH code to search for the best fit CMD, we find best fit values of D $=$ 3.31$\pm$0.07 Mpc and A$_{F475W}$ $=$ 0.05$\pm$0.04, both in excellent agreement with the independently measured values.  Examining the residual significance CMD, i.e., the difference between the data and the model weighted by the variance (Panel (d) of Figure \ref{fit}), we see a good, although not perfect fit.  Notably, the area between the blue helium burning stars and young MS appears to be too cleanly separated in the model, which could be due to differential extinction affecting young stars in the observed CMD.  Additionally, the model red helium burning stars are too blue compared to the data, likely due to uncertainties in the massive star models \citep[e.g.,][]{gza05}.  However, even the most discrepant regions are fit within $\pm$ 5$\sigma$, which are indicated by black or white points.  Overall, the model CMD appears to be in good agreement with the observed CMD, indicating that we have measured a reliable SFH.  See \citet{dol02} for a full discussion of the quality measures of this CMD fitting technique.  

In general, uncertainties in the absolute SFHs generally are somewhat anti-correlated between adjacent time bins, such that if the SFR is overestimated in one bin, it is underestimated in the adjacent bin.  However, cumulative SFHs,  i.e., the stellar mass formed during or previous to each time bin normalized to the integrated final stellar mass, do not share this property, and thus present a more robust way of analyzing SFHs. 

In the following sections, we consider SFHs plotted versus two different time binning schemes.  For the individual galaxies (Figure \ref{sfh1}), we present the cumulative SFHs at the highest time resolution possible.  Errors in the cumulative SFHs presented at this time resolution indicate both the uncertainty in the fraction of total stellar mass formed prior to a given time, and the inherent time resolution for the SFH of that particular galaxy.  A coarser time binning scheme is used for the absolute SFHs of the individual galaxies as well as the mean absolute and cumulative SFHs across the ANGST sample.  In Appendix \ref{appena}, we determine that six broad time bins of of 0-1, 1-2, 2-3, 3-6, 6-10, and 10-14 Gyr provide an optimal balance between photometric depths and age leverage contained in the CMDs of the ANGST dwarfs.  The broader time bins allow us to securely average the best fit SFHs from individual galaxies.  For the absolute SFHs of individual galaxies, broader time bins tend to minimize co-variance between adjacent time bins. Thus applying a broader time binning scheme to the absolute SFHs of individual galaxies provides for more secure measurements of the SFRs.

\begin{figure*}[h!]
\begin{center}
\epsscale{1.01}
\subfigure {
\plotone{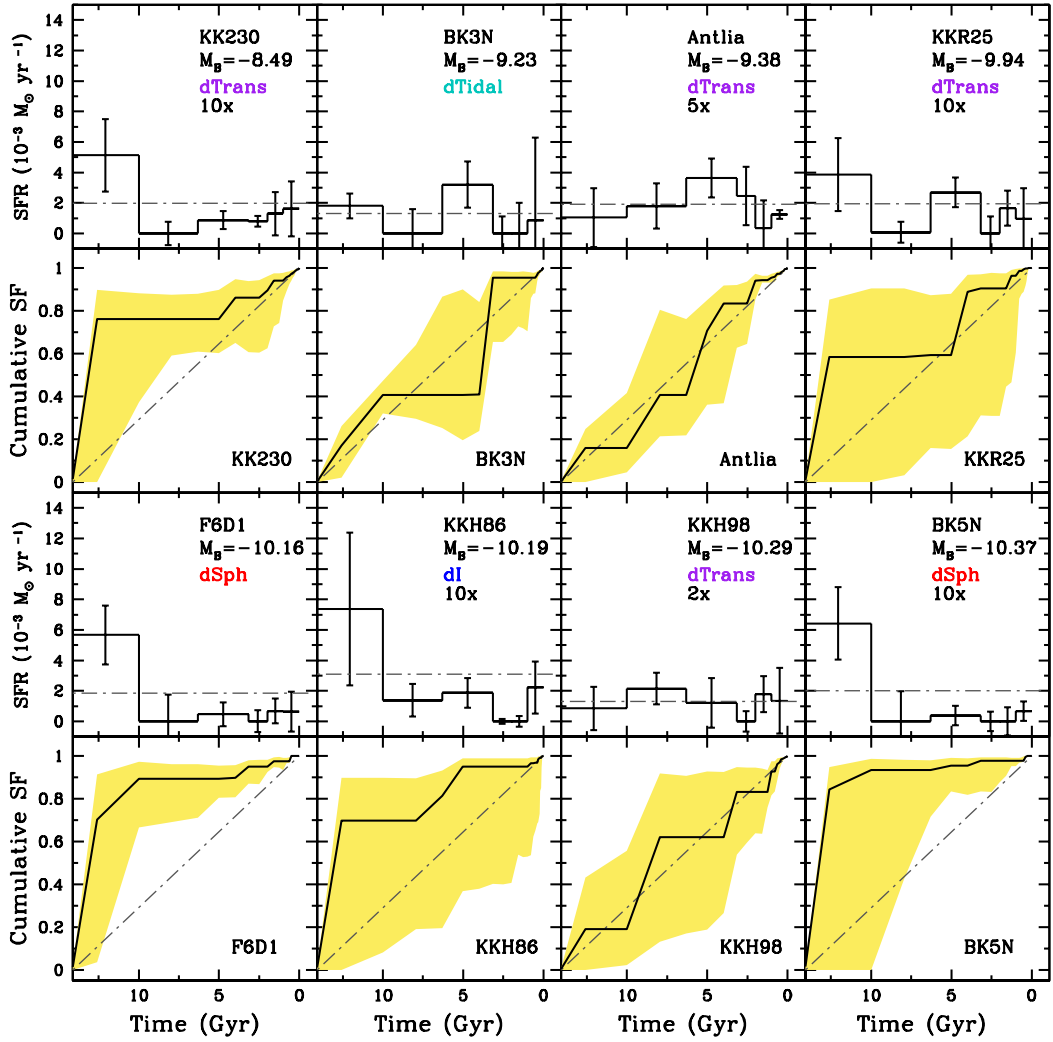}}
\caption{\small{The SFHs and cumulative SFHs, i.e., fraction of total stellar mass formed prior to or during a given time bin, of ANGST sample galaxies presented in order of increasing absolute blue luminosity.  The grey dashed line in the SFH plots is the lifetime averaged SFR.  In the cumulative SFHs, this same rate is also represented by the grey dashed line with a slope of unity, i.e., a constant SFH.  The error bars (or yellow envelope for the cumulative SFHs) shown are for the 16th and 84th percentile for the distribution of SFHs as computed via the Monte Carlo process described in \S \ref{sfhs}. The axes of the SFHs have been scaled so that galaxies of comparable luminosity are on similar scales.  Some galaxies absolute SFHs have been scaled up to clarify details.  Note that some of the SFRs have been scaled for clarity as indicated in the absolute SFH panel (e.g., `5x' means the SFH has been multiplied by a factor of 5 or `0.25x' means it has been scaled down by a factor of 4). The optical coverage fraction (see Table \ref{tab1}) indicates how much of the optical galaxy a SFH represents.}  }
\label{sfh1}
\end{center}
\end{figure*}

\addtocounter{figure}{-1}

\begin{figure*}[th!]
\addtocounter{subfigure}{-1}
\begin{center}
\epsscale{1.01}
\subfigure {
\plotone{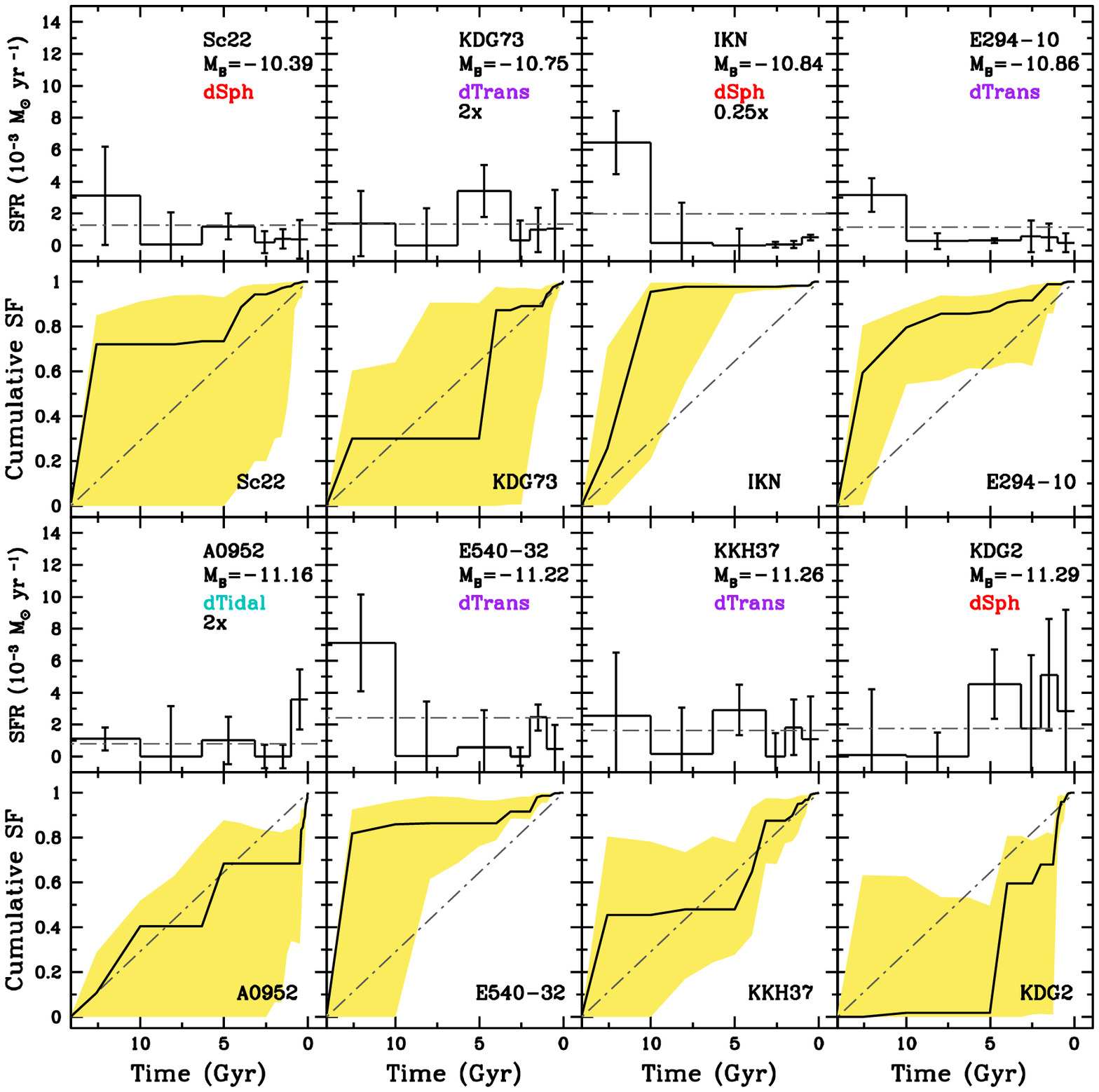}}
\caption{Continued}  
\label{sfh2}
\end{center}
\end{figure*}

\addtocounter{figure}{-1}

\begin{figure*}[th!]
\addtocounter{subfigure}{-1}
\begin{center}
\subfigure {
\plotone{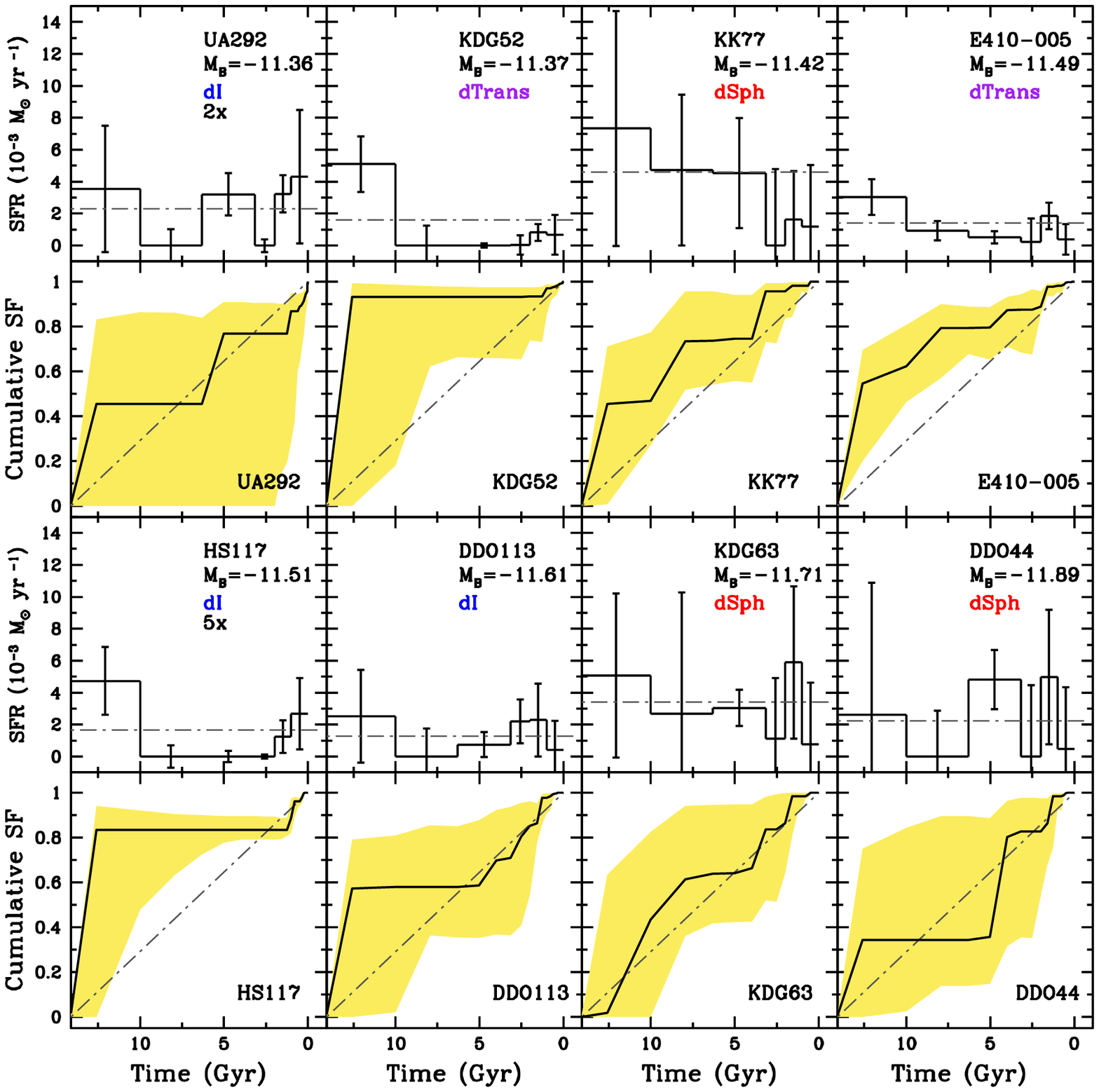}}
\caption{Continued}  
\label{sfh3}
\end{center}
\end{figure*}

\addtocounter{figure}{-1}

\begin{figure*}[ht!]
\addtocounter{subfigure}{-1}
\begin{center}
\subfigure {
\plotone{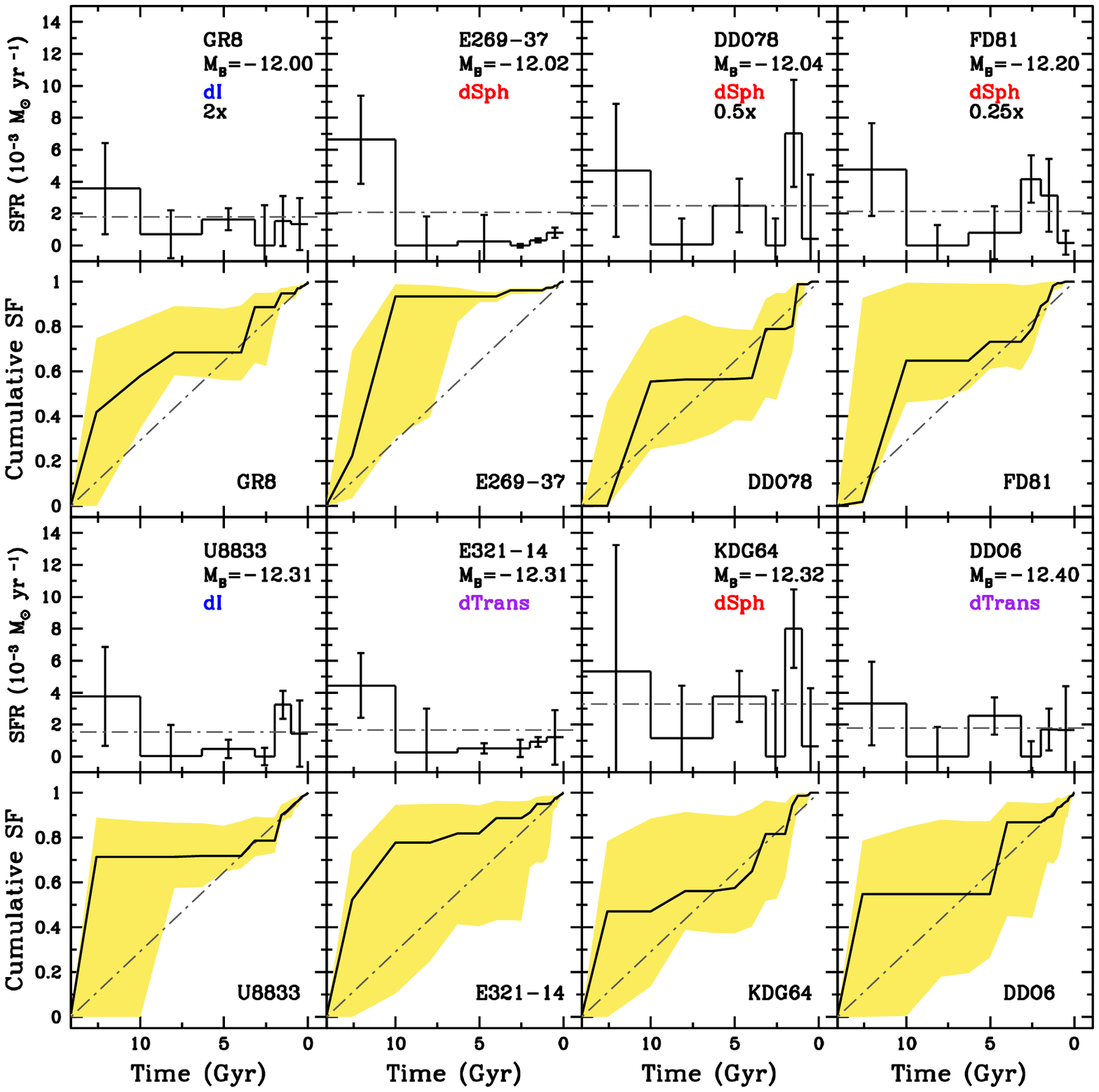}}
\caption{Continued}  
\label{sfh4}
\end{center}
\end{figure*}

\addtocounter{figure}{-1}

\begin{figure*}[ht!]
\addtocounter{subfigure}{-1}
\begin{center}
\subfigure {
\plotone{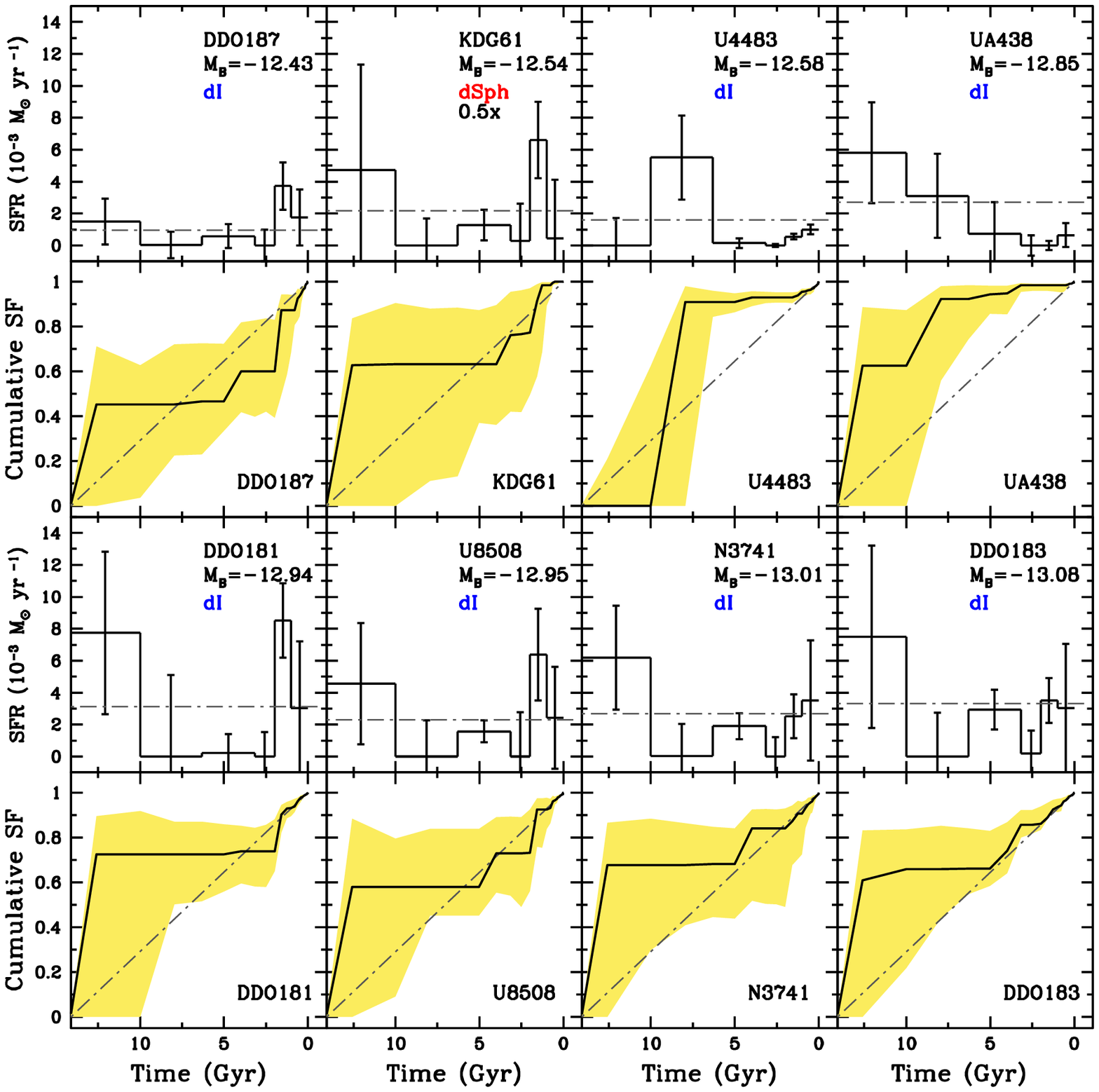}}
\caption{Continued}  
\label{sfh5}
\end{center}
\end{figure*}

\addtocounter{figure}{-1}

\begin{figure*}[ht!]
\addtocounter{subfigure}{-1}
\begin{center}
\subfigure {
\plotone{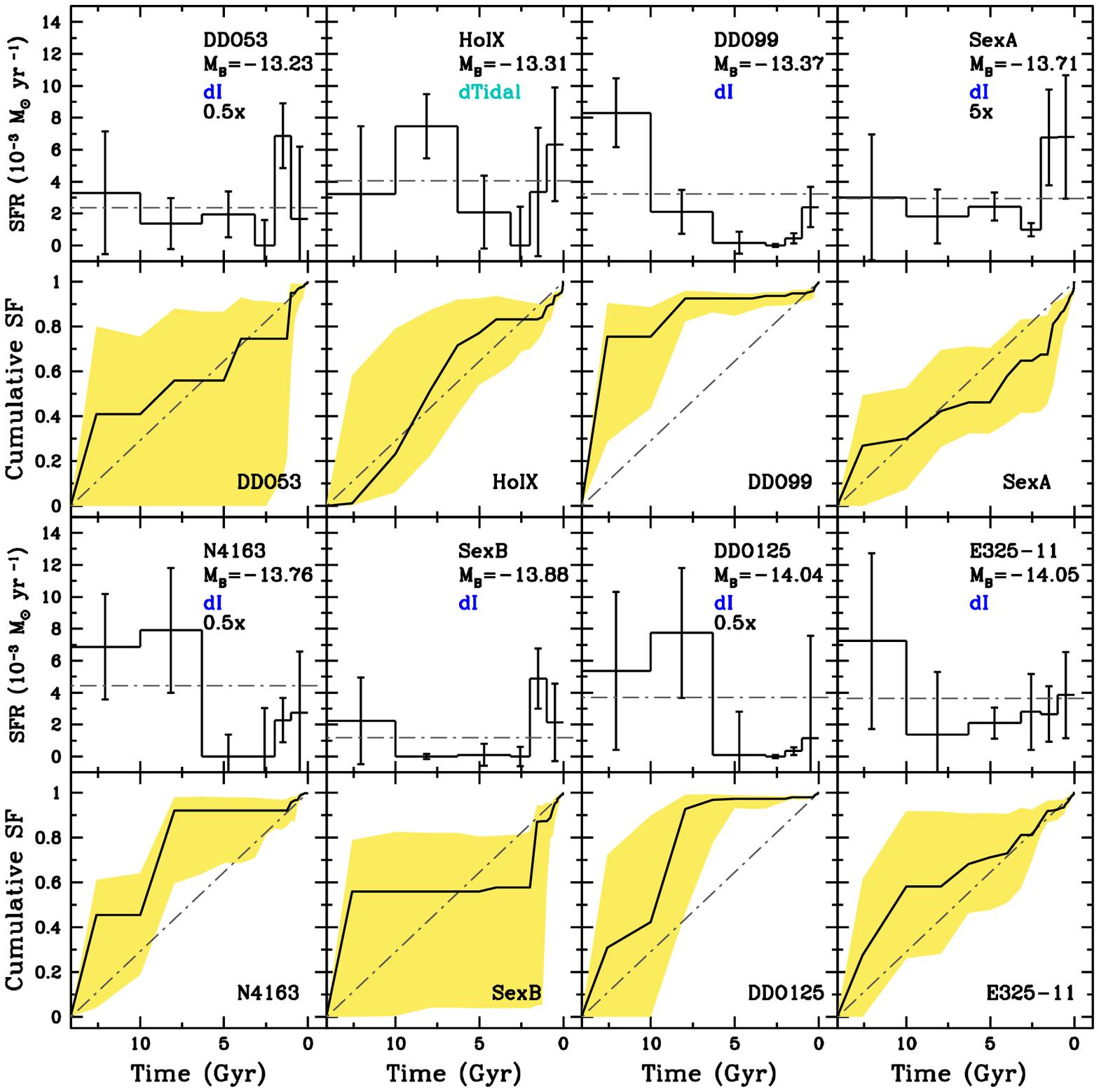}}
\caption{Continued}  
\label{sfh6}
\end{center}
\end{figure*}

\addtocounter{figure}{-1}

\begin{figure*}[ht!]
\addtocounter{subfigure}{-1}
\begin{center}
\subfigure {
\plotone{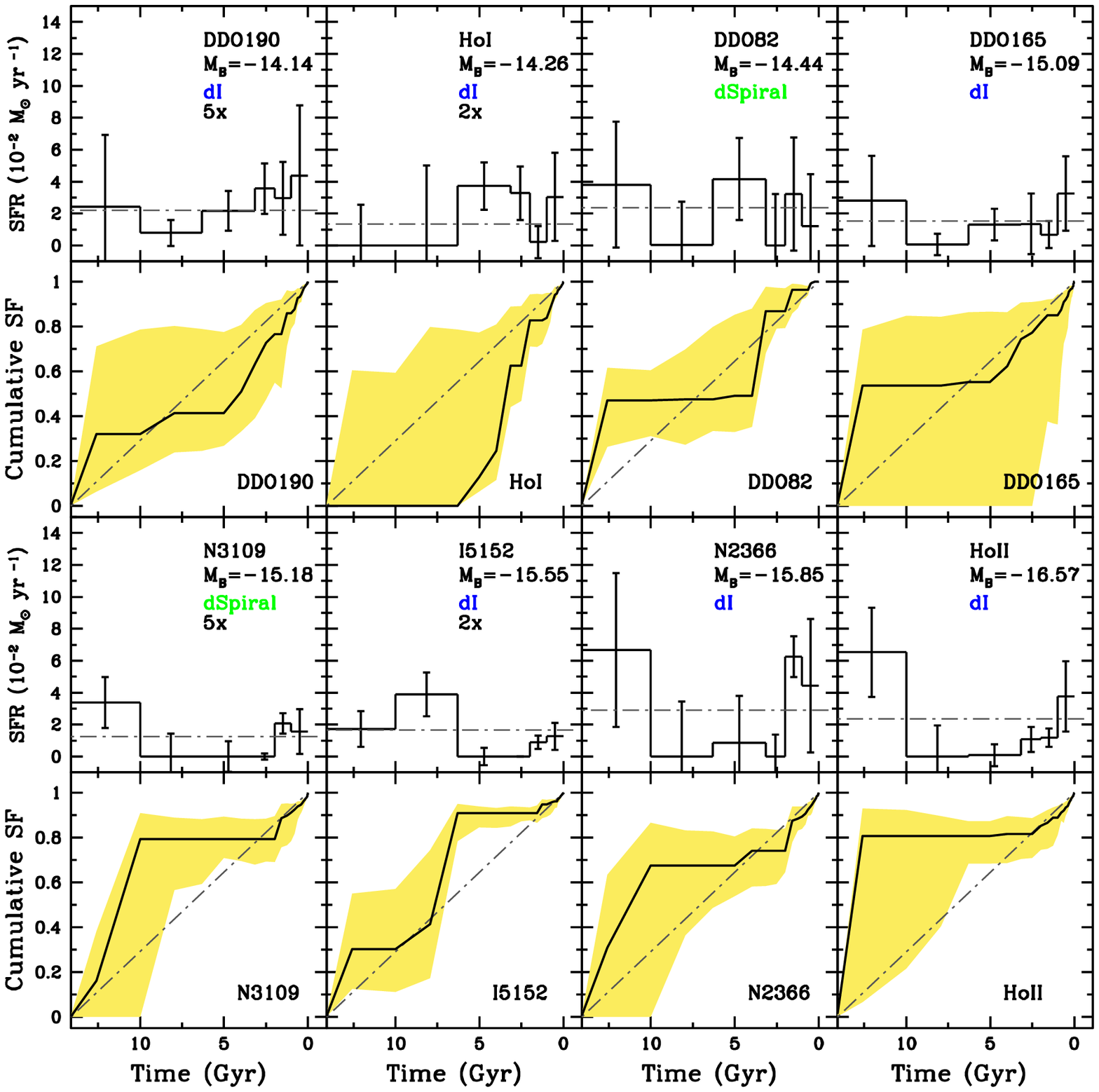}}
\caption{Continued}  
\label{sfh7}
\end{center}
\end{figure*}

\addtocounter{figure}{-1}

\begin{figure*}[h!]
\addtocounter{subfigure}{-1}
\begin{center}
\subfigure {
\plotone{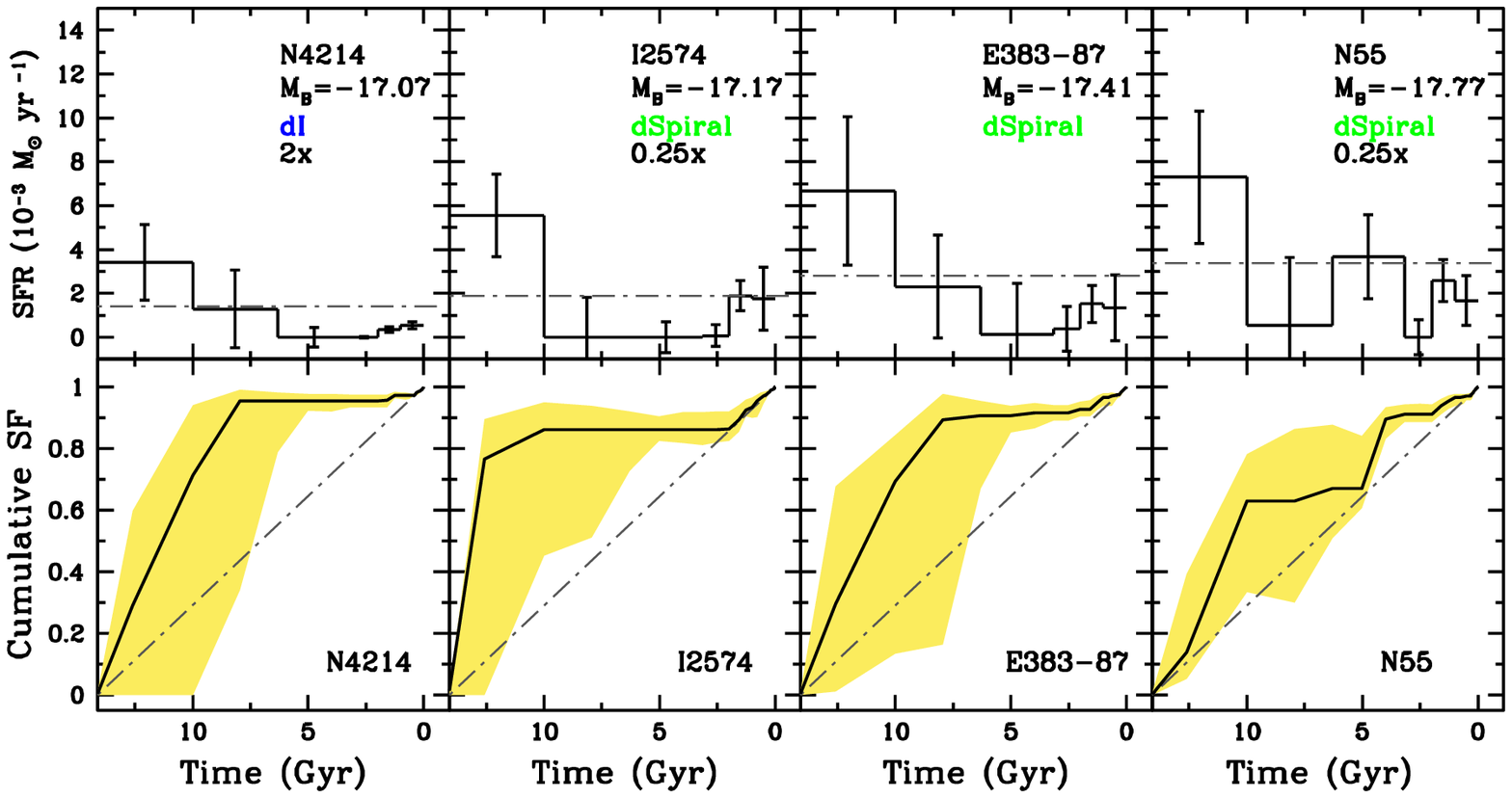}}
\caption{Continued}  
\label{sfh8}
\end{center}
\end{figure*}

\begin{deluxetable*}{lccccccccc}
\tablecolumns{10}
\tabletypesize{\footnotesize}
\tablewidth{0pt}
\tablecaption{SF Properties of the ANGST Dwarf Galaxies}

\tablehead{
    \colhead{Galaxy} &
    \colhead{$\langle$Lifetime SFR$\rangle$} &
        \colhead{$M_{25}$} &
         \colhead{$M_{25}$/$L_{B}$} &
   \colhead{$f_{10Gyr}$} &
    \colhead{$f_{6Gyr}$} &
       \colhead{$f_{3Gyr}$} &
       \colhead{$f_{2Gyr}$} &
       \colhead{$f_{1Gyr}$} &
        \colhead{$f_{Current}$}\\
         \colhead{} &
    \colhead{(10$^{-3}$ $M_{\odot} yr^{-1}$) } &
     \colhead{(10$^{7}$ $M_{\odot}$) } &
          \colhead{} &
          \colhead{} &
      \colhead{} &
      \colhead{} &
      \colhead{} &
     \colhead{} &
      \colhead{}  \\
          \colhead{(1)} &
    \colhead{(2)} &
        \colhead{(3)} &
         \colhead{(4)} &
    \colhead{(5)} &
       \colhead{(6)} &
      \colhead{(7)} &
      \colhead{(8)} &
      \colhead{(9)} &
      \colhead{(10)}
      }

\startdata

KK230 & 	0.20 & 0.006 & 0.15 &  0.76 & 0.76 & 0.86 & 0.89 & 0.94 & 1.00 \\
BK3N & 	1.40 & 0.03 & 0.37 & 0.41 & 0.41 & 0.95 & 0.95 & 0.95 & 1.00 \\
Antlia & 	0.38 & 0.10 & 1.30 & 0.16 & 0.41 & 0.83 & 0.94 & 0.95 & 1.00\\
KKR25 & 	0.19 & 0.03 & 0.19 & 0.58 & 0.59 & 0.90 & 0.90 & 0.97 & 1.00\\
FM1 & 	1.8 & 0.13 & 0.69 & 0.89 & 0.89 & 0.95 & 0.95 & 0.97 & 1.00\\
KKH86 & 	0.31 & 0.02 & 0.12 & 0.70 & 0.81 & 0.94 & 0.94 & 0.94  & 1.00\\
KKH98 & 	0.64 & 0.41 &  0.20 & 0.19 & 0.62 & 0.83 & 0.83 & 0.93 & 1.00\\
BK5N & 	2.0 & 0.10 & 0.46 & 0.93 & 0.93 & 0.97 & 0.97 & 0.97 & 1.00\\
Sc22 & 	1.26 & 0.08 & 0.35 & 0.72 & 0.73 & 0.94 & 0.96 & 0.98 & 1.00\\
KDG73 & 	0.65 & 0.02 &  0.06 & 0.30 & 0.30 & 0.87 & 0.89 & 0.94 & 1.00\\
IKN & 	7.9 & 4.8 & 14.0 & 0.02 & 0.95 & 0.98 & 0.98 & 9.98 & 1.00\\
E294-010 & 	1.16 & 0.09 & 0.26 &  0.80 & 8.86 & 0.92 & 0.96 & 0.99 & 1.00\\
A0952+69 & 	0.46 & 0.13 & 0.29 & 0.40 & 0.40 & 0.68 & 0.68 & 0.68 & 1.00\\
E540-032 & 	0.24 & 0.36 & 0.76 & 0.86 & 0.86 & 0.91 & 0.91 & 0.98 & 1.00\\
KKH37 & 	1.64 & 0.16 & 0.31 &  0.45 & 0.48 & 0.87 & 0.87 & 0.95 & 1.00\\
KDG2 & 	0.34 & 0.04 & 0.09 & 0.02 & 0.02 & 0.60 & 0.68 & 0.88 & 1.00\\
UA292 & 	0.47 & 0.03 & 0.06 & 0.45 & 0.45 & 0.77 & 0.77 & 0.87 & 1.00\\
KDG52 & 	1.60 & 0.24 & 0.44 & 0.93 & 0.93 & 0.93 & 0.94 & 0.97 & 1.00\\
KK77 & 	4.6 & 1.94 & 3.4 & 0.47 & 0.74 & 0.96 & 0.96 & 0.98 & 1.00\\
E410-005 & 1.42 & 0.18 & 0.30 & 0.62 & 0.79 & 0.87 & 0.89 & 0.98 & 1.00\\
HS117 & 	0.33 &  0.04 & 0.07 &0.83 & 0.83 & 0.83 & 0.83 & 0.89 & 1.00\\
DDO113 & 1.30 & 0.24 & 0.36 & 0.58 & 0.58 & 0.71 & 0.85 & 0.98 & 1.00\\
KDG63 & 	3.4 & 0.85 & 1.10 & 0.43 & 0.64 & 0.84 & 0.86 & 0.98 & 1.00\\
DDO44 & 	2.20 & 1.30 & 1.5 & 0.34 & 0.34 & 0.83 & 0.83 & 0.98 & 1.00\\
GR8 & 	0.90 & 0.10 & 0.1 & 0.59 & 0.69 & 0.88 & 0.88 & 0.95 & 1.00\\
E269-37  & 2.00 & 0.20 & 0.2 & 0.94 & 0.94 & 0.96 & 0.96 & 0.97 & 1.00\\
DDO78 & 	4.75 & 1.77 & 1.7 & 0.56 & 0.56 & 0.79 & 0.79 & 0.99 & 1.00\\
F8D1 & 	8.35 & 3.8 & 3.2 & 0.65 & 0.65 & 0.73 & 0.89 & 0.99 & 1.00\\
U8833 & 	1.54 & 0.11 & 0.08 & 0.71 & 0.72 & 0.79 & 0.79 & 0.94 & 1.00\\
E321-014 & 1.66 & 0.29 & 0.22 & 0.77 & 0.82 & 0.89 & 0.92 & 0.95 & 1.00\\
KDG64 & 	3.35 & 0.56 & 0.42 & 0.47 & 0.56 & 0.82 & 0.82 & 0.98 & 1.00\\
DDO6 & 	1.74 & 0.20 & 0.14 & 0.55 & 0.55 & 0.87 & 0.87 & 0.93 & 1.00\\
DDO187 & 0.97 & 0.21 & 0.14 & 0.45 & 0.47 & 0.60 & 0.60 & 0.87 & 1.00\\
KDG61 & 	4.35 & 1.4 & 0.88 & 0.63 & 0.63 & 0.76 & 0.77 & 0.98 & 1.00\\
U4483 & 	1.57 & 0.27 & 0.16 &0.00 & 0.91 & 0.93 & 0.93 & 0.96 & 1.00\\
UA438 & 	2.74 & 1.0 & 0.48 & 0.63 & 0.92 & 0.98 & 0.98 & 0.98 & 1.00\\
DDO181 & 3.11 & 0.92 & 0.39 & 0.72 & 0.72 & 0.74 & 0.74 & 0.93 & 1.00\\
U8508 & 	2.33 & 0.39 &  0.16 & 0.58 & 0.58 & 0.73 & 0.73 & 0.93 & 1.00\\
N3741 & 	2.63 & 0.57 & 0.23 & 0.68 & 0.68 & 0.84 & 0.84 & 0.91 & 1.00\\
DDO183 & 3.31 & 0.50 & 0.19 & 0.66 & 0.66 & 0.86 & 0.86 & 0.94 & 1.00\\
DDO53 & 	4.75 & 1.0 & 0.34 & 0.41 & 0.56  &0.75 & 0.75 & 0.95 & 1.00\\
\hoix\ & 	4.04 & 1.15 & 0.35 & 0.23 & 0.72 & 0.83 & 0.83 & 0.89 & 1.00\\
DDO99 & 	3.12 & 4.0 & 1.20 & 0.75 & 0.93 & 0. 94 & 0.94 & 0.95 & 1.00\\
SexA & 	0.59 & 3.53 & 0.75 & 0.30 & 0.46 & 0.65 & 0.67 & 0.84 & 1.00\\
N4163 & 	8.93 & 2.65 & 0.54 & 0.45 & 0.92 & 0.92 & 0.92 & 0.96 & 1.00\\
SexB & 	1.17 & 4.38 & 0.78 & 0.56 & 0.56 & 0.58 & 0.58 & 0.87 & 1.00\\
DDO125 & 7.26 & 15.6 & 2.40 & 0.42 & 0.97 & 0.97 & 0.97 & 0.98 & 1.00\\
E325-11 & 3.59 & 2.61 & 0.40 & 0.58 & 0.68 & 0.81 & 0.87 & 0.93 & 1.00\\
DDO190 & 4.29 & 1.21 & 0.17 & 0.32 & 0.41 & 0.63 & 0.77 & 0.86 & 1.00\\
\hoi\ & 	6.65 & 6.83 & 0.87 &0.00 & 0.00 & 0.62 & 0.82 & 0.84 & 1.00\\
DDO82 & 	25.1 & 16.8 & 1.80 & 0.47 & 0.48 & 0.87 & 0.87 & 0.96 &  1.00\\
DDO165 & 15.1 & 9.77  & 0.58 & 0.54 & 0.55 & 0.74 & 0.82 & 0.85 & 1.00\\
N3109 & 	2.64 & 20.2 & 1.10 & 0.79 & 0.79 & 0.79 & 0.79 & 0.91 & 1.00\\
I5152 & 	8.50 & 29.6 & 1.10 & 0.30 & 0.91 & 0.01 & 0.91 & 0.95 & 1.00\\ 
N2366 & 	28.5 &  26.8 & 0.79 & 0.67 & 0.67 & 0.74 & 0.74 & 0.89 & 1.00\\
\hoii\ & 	23.4 & 61.6 & 0.93 & 0.81 & 0.81 & 0.82 & 0.85 & 0.89 & 1.00\\
N4214 & 	7.02 & 82.3 & 0.79 & 0.71 & 0.95 & 0.95 & 0.95 & 0.97 & 1.00\\
I2574 & 	74.3 & 175 & 1.50 & 0.86 & 0.86 & 0.86 & 0.86 & 0.93 & 1.00\\
E383-87 & 28.5 & 168 & 1.90 & 0.69 & 0.91 & 0.92 & 0.93 & 0.97 & 1.00\\
N55 & 	138.5 & 1210 & 6.10 & 0.63 & 0.67 & 0.91 & 0.91 & 0.97 & 1.00
\enddata
\tablecomments{\scriptsize{SF properties of the ANGST dwarf galaxies from the CMD-based SFHs -- (1) Galaxy Name; (2) Lifetime averaged SFR, i.e., average SFR over the history of the galaxy, from the best fit SFH; (3) Integrated stellar mass from the best fit SFH, with the $\mathcal{A}_{25}$ normalization applied; (4) Indicative mass-to-light ratio in solar units; (5)--(10) The cumulative fraction of stars formed prior to 10, 6, 3, 2, 1 Gyr ago from the best fit SFHs.}}
\label{tab2}
\end{deluxetable*}
%
\clearpage

\section{The Extended Star Formation Histories of Dwarf Galaxies} 
\label{results}

The ANGST dwarf galaxies exhibit a wide variety of complex SFHs.  In Figure \ref{sfh1} we show the absolute SFHs, i.e., SFR($t$), and the cumulative SFHs of the individual galaxies, sorted in order of increasing blue luminosity.  A cursory inspection of the 60 SFHs reveals that often galaxies with similar luminosities, morphologies, or chemical compositions, do not have consistent SFHs, confirming the complexity of dwarf galaxy SFHs previously found in studies of the LG \citep[e.g.,][]{mat98, dol05, tol09}. The wide variety of SFHs underlines the importance of having a large sample; conclusions based on the SFH of a single galaxy may not be be representative of the population as a whole.

This paper leverages the size of the ANGST sample to explore general trends seen in dwarf galaxies, e.g., differences among the morphological types.  While the SFHs of individual galaxies can provide useful insight into a particular system's evolutionary state, we generally do not consider individual galaxies in this paper.  However, in Table \ref{tab2} we provide data on the SFHs of individual galaxies considered in this sample, which may be useful for specific studies.

\begin{deluxetable*}{lcccccccc}
\tablecolumns{8}
\tabletypesize{\footnotesize}
\tablewidth{0pt}
\tablecaption{Mean Star Formation Properties per Morphological Type}
\tablehead{ 
\colhead{Group} &  
\colhead{$\langle$$M_{25}$$\rangle$} & 
    \colhead{$f_{10Gyr}$} &
    \colhead{$f_{6Gyr}$} &
       \colhead{$f_{3Gyr}$} &
       \colhead{$f_{2Gyr}$} &
       \colhead{$f_{1Gyr}$} &
        \colhead{$f_{Current}$} \\      
        \colhead{} &
    \colhead{(10$^{7}$ \msun)} &
    \colhead{} &
        \colhead{} &
       \colhead{} &
       \colhead{} &
       \colhead{}  &
      \colhead{}  \\
           \colhead{(1)} &
    \colhead{(2)} &
    \colhead{(3)} &
        \colhead{(4)} &
       \colhead{(5)} &
       \colhead{(6)} &
       \colhead{(7)}  &
      \colhead{(8)}   
}    
\startdata  
dSph & 5.4$\pm$0.9 & 0.65$^{+0.06}_{-0.08}$ & 0.71$^{+0.05}_{-0.06}$& 0.87$^{+0.03}_{-0.03}$  & 0.89$^{+0.03}_{-0.02}$ & 0.98$^{+0.00}_{-0.00}$ & 1.00\\
dI & 7.3$\pm$1.8 &  0.54$^{+0.05}_{-0.04}$ & 0.67$^{+0.03}_{-0.04}$& 0.80$^{+0.03}_{-0.02}$  & 0.82$^{+0.02}_{-0.01}$ & 0.92$^{+0.01}_{-0.01}$ & 1.00\\
dTrans & 2.8$\pm$1.5 & 0.53$^{+0.10}_{-0.09}$ & 0.63$^{+0.09}_{-0.08}$& 0.86$^{+0.01}_{-0.02}$  & 0.89$^{+0.01}_{-0.02}$ & 0.96$^{+0.01}_{-0.01}$ & 1.00\\
dSpiral & 74$\pm$35 & 0.59$^{+0.14}_{-0.11}$ & 0.71$^{+0.10}_{-0.06}$& 0.88$^{+0.03}_{-0.02}$  & 0.89$^{+0.03}_{-0.02}$ & 0.95$^{+0.01}_{-0.01}$ & 1.00\\
dTidal & 1.6$\pm$0.35 & 0.38$^{+0.08}_{-0.05}$ & 0.64$^{+0.09}_{-0.05}$& 0.83$^{+0.06}_{-0.07}$  & 0.84$^{+0.07}_{-0.07}$ & 0.85$^{+0.08}_{-0.06}$ & 1.00
\enddata
\tablecomments{Mean SF properties for the ANGST dwarf galaxies, grouped by morphological type.  (2) The integrated stellar mass normalized to the $\mathcal{A}_{25}$ areal fraction and then averaged per morphological type; (3)--(7) The cumulative fraction of total stellar mass formed prior to 10, 6, 3, 2, 1 Gyr ago.  Error bars represent the uncertainty in the mean value.}
\label{tab3}
\end{deluxetable*}

\begin{figure}[th!]
\begin{center}
\epsscale{1.25}
\plotone{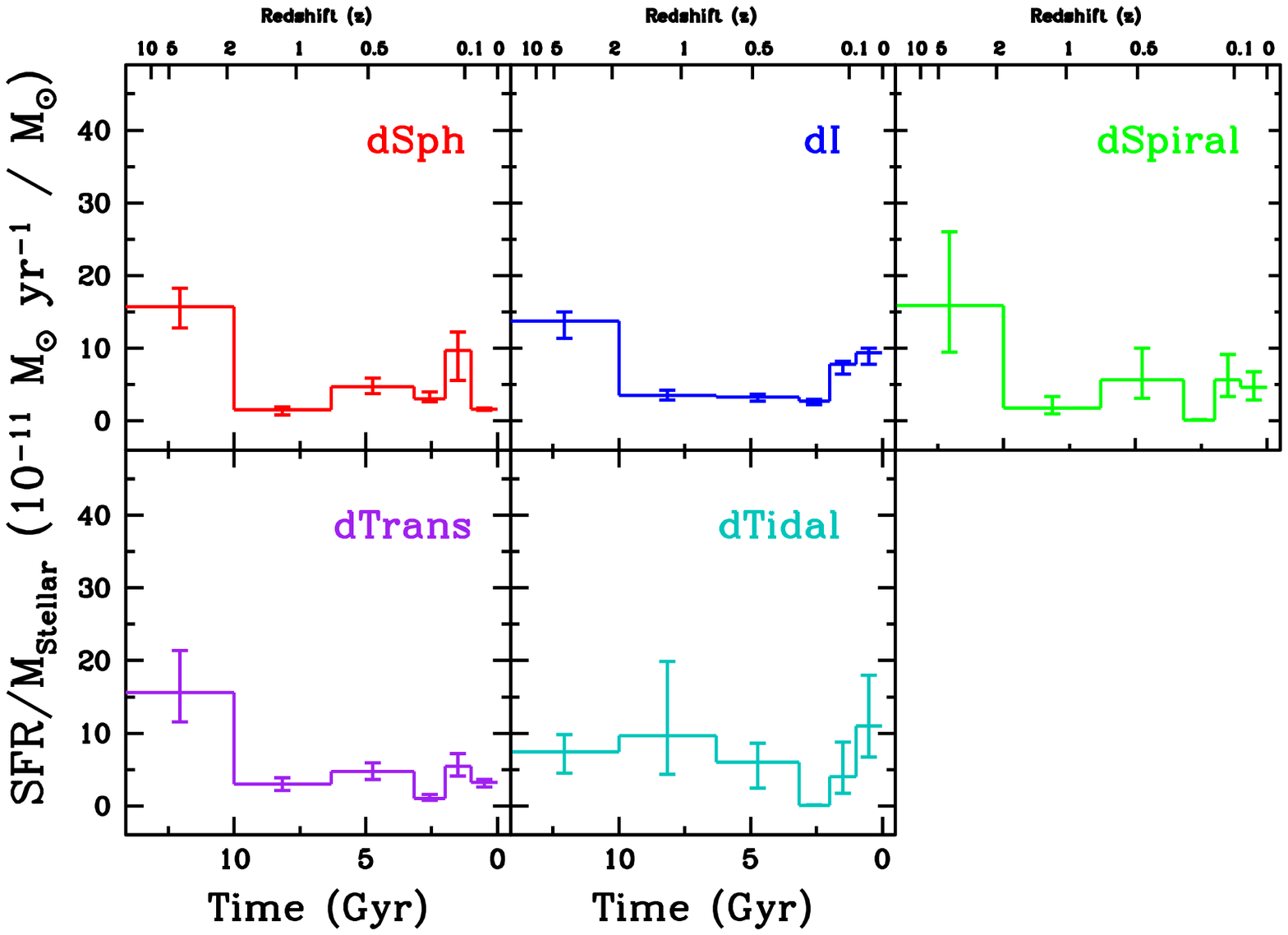}
\caption{The mean specific SFHs (sSFH), i.e., the SFH divided by the integrated stellar mass, for each morphological type.  The error bars reflect the uncertainties in the mean sSFHs.  Generally, dwarf galaxies had a high level of SF at ancient times ($>$ 10 Gyr ago), with a low level of SF at intermediate times (1--10 Gyr ago). There are noticeable differences in the sSFHs among the morphological types within the most recent 1 Gyr.}  
\label{sfh_avg}
\end{center}
\end{figure}

We first consider the unweighted mean specific SFHs (i.e., a SFH divided by the total integrated stellar mass formed) of the ANGST sample grouped by morphological type (Figure \ref{sfh_avg}).  Comparison among the specific SFHs shows general consistency for times $\gtrsim$ 1 Gyr ago, with the exception of dTidals (note that dTidals will be discussed in greater detail in the context of recent SFHs in \citealt{weiprep}).  Qualitatively, we find that a typical dwarf galaxy exhibits dominant ancient SF ($>$ 10 Gyr ago), and lower levels of SF at intermediate times (1--10 Gyr ago).  The only consistent difference among the morphological types is within the last 1 Gyr, where dSphs exhibit a significant drop in SFRs relative to the other types.  This suggests that many of the dSphs in the ANGST sample could have been gas-rich as recently as 1 Gyr ago, implying that the process of gas loss can occur relatively quickly and at late times.  It is tempting to associate the sharp drop in the SFR of a typical dSph $\sim$ 1 Gyr ago with rapid gas loss. However, broad uncertainties in the AGB models \citep[e.g.,][]{gir10, mar10} make the precise age for the drop in dSph SFRs and the degree of synchronization uncertain.  In addition, subsamples of the dSphs do appear to have dramatically lower SFRs at more intermediate ages.

\begin{figure}[]
\begin{center}
\epsscale{1.2}
\plotone{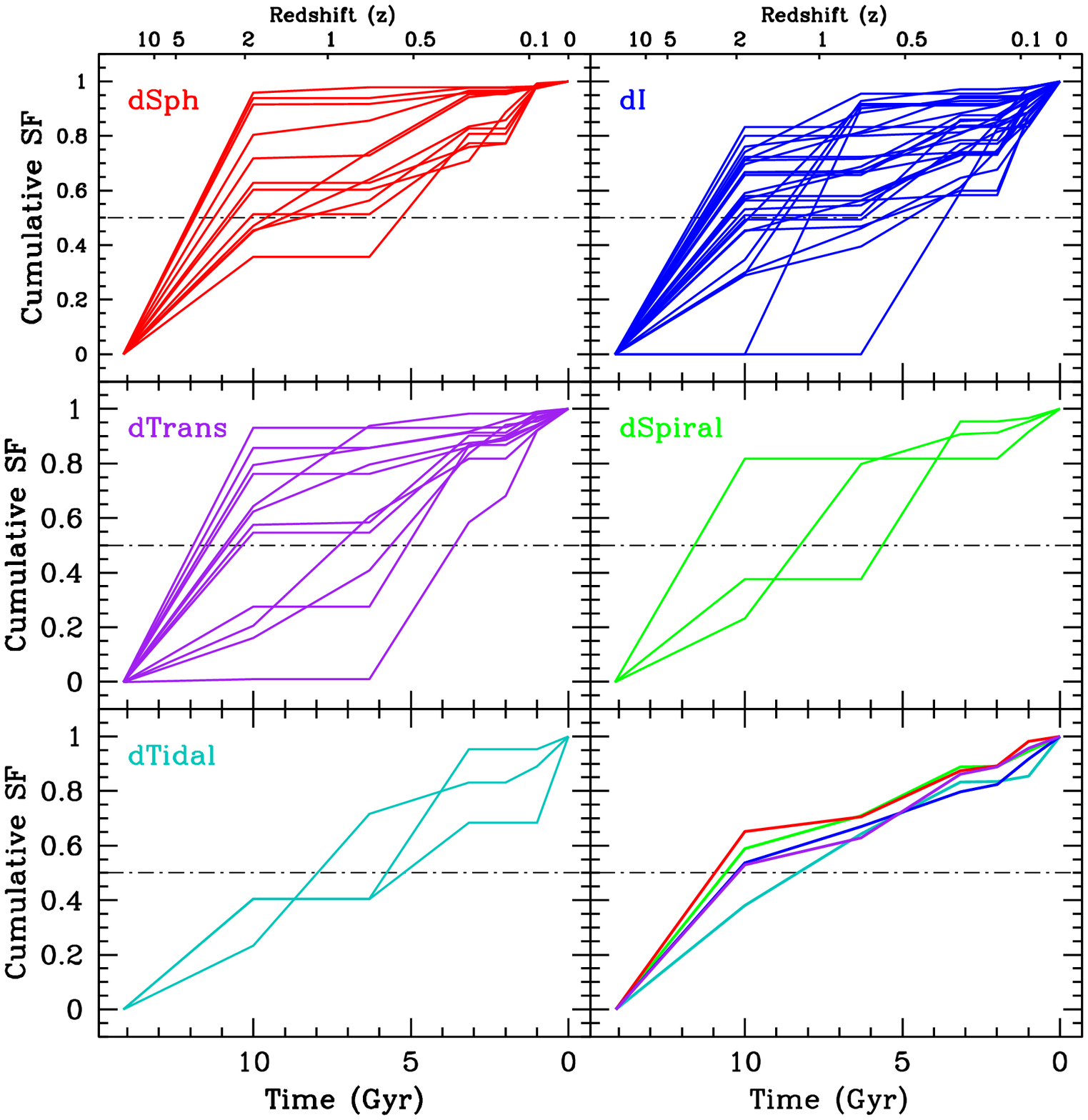}
\caption{The individual and mean cumulative SFHs, per morphological type.  Error bars have been omitted for clarity, and are instead listed in Tables \ref{tab2} and \ref{tab3}.  The dot-dashed line represents 50\% of the total stellar mass. While the individual cumulative SFHs show a wide variety, the mean cumulative SFHs are remarkably similar. Most dwarf galaxies are not entirely old stellar populations, i.e., they have intermediate or recent SF, and none are consistent with simple models of SF, i.e., single epoch or constant SFHs. Excluding dTidals, the average dwarf galaxy formed $\gtrsim$ 50\% of its stars prior to z $\sim$ 2 (10 Gyr ago).}  
\label{cum1}
\end{center}
\end{figure}

Cumulative SFHs allow us to readily compare galaxies of different masses. Like the absolute SFHs, the cumulative SFHs plotted in Figure \ref{cum1} show significant diversity within each morphological class, yet converge on mean values that are broadly consistent among the morphological types.  Perhaps the most striking result is that the average dwarf galaxy formed $\gtrsim$ 50\% of its total stellar mass by z $\sim$ 2 (10 Gyr ago) and $\gtrsim$ 60\% by z $\sim$ 1 (7.6 Gyr ago), independent of morphological type (Table \ref{tab3}). 

Although dSphs appear to have formed a slightly larger percentage of total stellar mass than dIs by z $\sim$ 2, additional consideration of systematic uncertainties, as discussed in Appendix \ref{systematics}, indicate that the ANGST data is not deep enough to make such fine distinctions.  More importantly, the amplitude of the difference is significantly smaller than typical models, which assume that dSphs are dominated by ancient stellar populations and dIs are consistent with constant SF over their lifetimes  \citep[e.g.,][]{bin94, ski95, gre03}.

\begin{figure*}[t]
\begin{center}
\epsscale{0.9}
\plotone{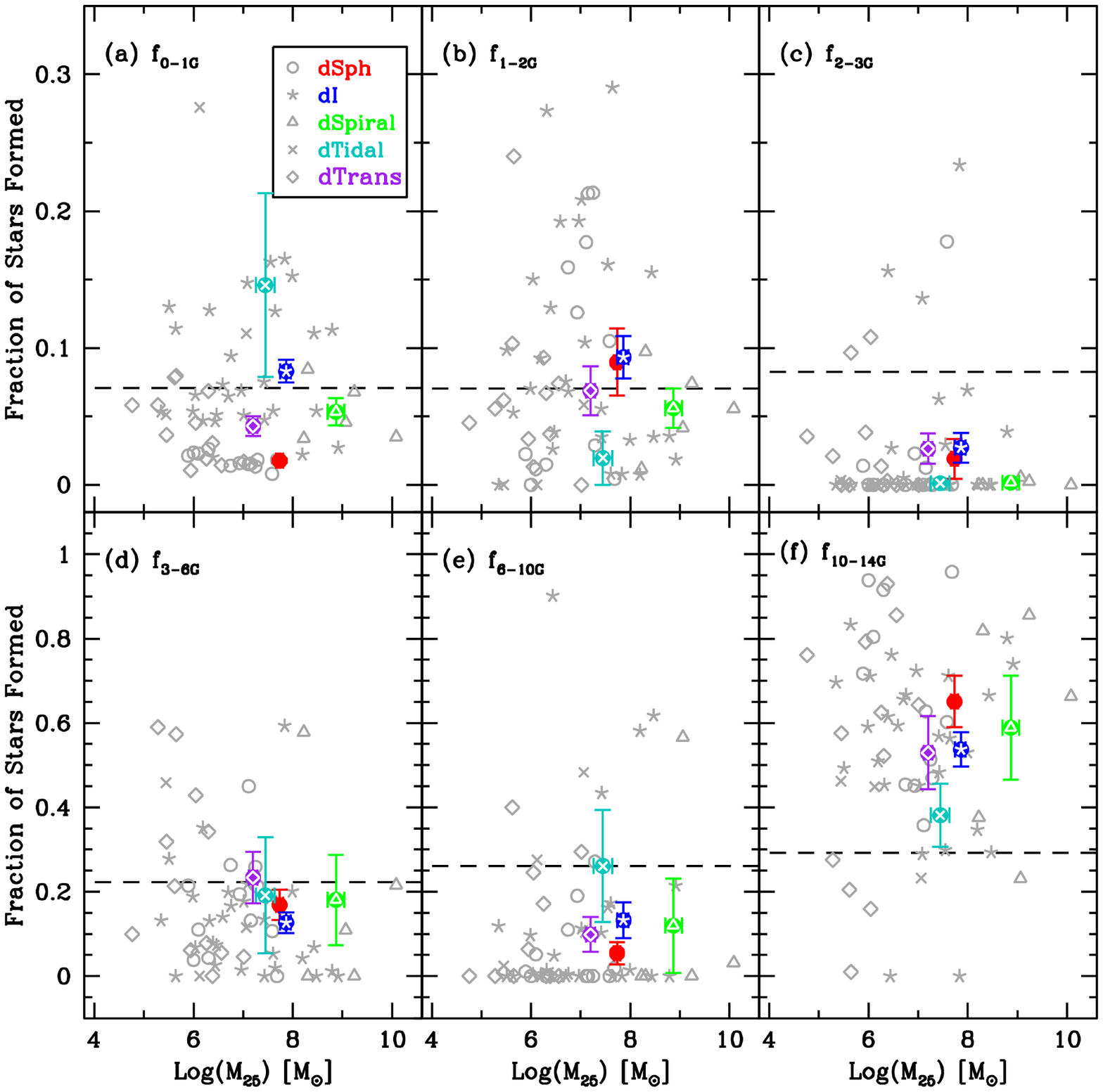}
\caption{The fractional SFHs, i.e., the fraction of total stellar mass formed during a given time bin, for each galaxy shown as a function of the total stellar mass and morphological type. $M_{25}$ is derived by integrating the SFHs (Figure \ref{sfh1}), and normalizing to $\mathcal{A}_{25}$ to account for differences in the observed areas. The time bins are: $f_{0-1G}$ $=$ 0-1, $f_{1-2G}$ $=$ 1-2, $f_{2-3G}$ $=$ 2-3, $f_{3-6G}$ $=$ 3-6, $f_{6-10G}$ $=$ 6-10, $f_{10-14G}$ $=$ 10-14 Gyr ago. The grey points represent individual galaxies, while the colored points show the mean values per morphological type with error bars representing the uncertainty in the mean.}
\label{mbf_avg}
\end{center}
\end{figure*}

The mean cumulative SFHs hint at divergence among the morphological types within the last few Gyr (Figure \ref{cum1}).   However, the exact characteristics of differences in the SFHs during this time period are ambiguous. Broad uncertainties in the AGB star models \citep[e.g.,][]{gir10, mar10}, our modest time resolution, and systemic effects on the mean cumulative SFH (see Appendix \ref{systematics}) all affect our ability to quantitatively examine the details of intermediate age SF.

We do identify secure measurements of differences within the most recent 1 Gyr.  At these times,  luminous MS and core blue and red helium burning stars provide excellent age leverage \citep[e.g.,][]{doh98}, and the SFHs clearly illustrate differences among the morphological types.  Specifically, the typical dSph, dI, dTrans, and dSpiral formed $\sim$ 2\%, 8\%, 4\%, and 5\% of their total stellar mass within the most recent 1 Gyr.  The combined findings from intermediate and recent SFHs suggest that morphological differences and complete gas loss in dwarf galaxies can be relatively recent phenomena, at least within the Local Volume.  dSphs that are strictly old, i.e., have no AGB star populations, are known to exist in the LG \citep[e.g.,][]{mou82, mat98, dol05, tol09}, but appear to only represent a minority of all known dSphs in the larger volume we analyze here.  This is at least in part due to the detection limit of the ANGST sample as discussed in \S \ref{selection}.

To further explore the differences among the morphological types, we consider the fractional SFH, i.e., the fraction of stellar mass formed in each time bin, as a function of the total normalized stellar mass, $M_{25}$ (computed by integrating the SFH over time and applying the $\mathcal{A}_{25}$ area normalization).  We show values for both the individual galaxies (grey symbols) and the mean values per morphological type, with error bars representative of the uncertainty in the mean in Figure \ref{mbf_avg}.  The mean stellar masses and cumulative SFHs for each of the morphological types are listed in Table \ref{tab3}.  

The mean stellar masses show that dSpirals are typically the most massive galaxies, dTrans are the least massive, and dIs and dSphs have similar mean masses.  There are not strong links between patterns of SF, total stellar mass, and morphological type, confirming that SF processes among the different types of dwarf galaxies are not dramatically different.  We continue analysis involving the integrated stellar masses in \S \ref{gascomp}.

\begin{figure*}[t]
\begin{center}
\epsscale{0.9}
\plotone{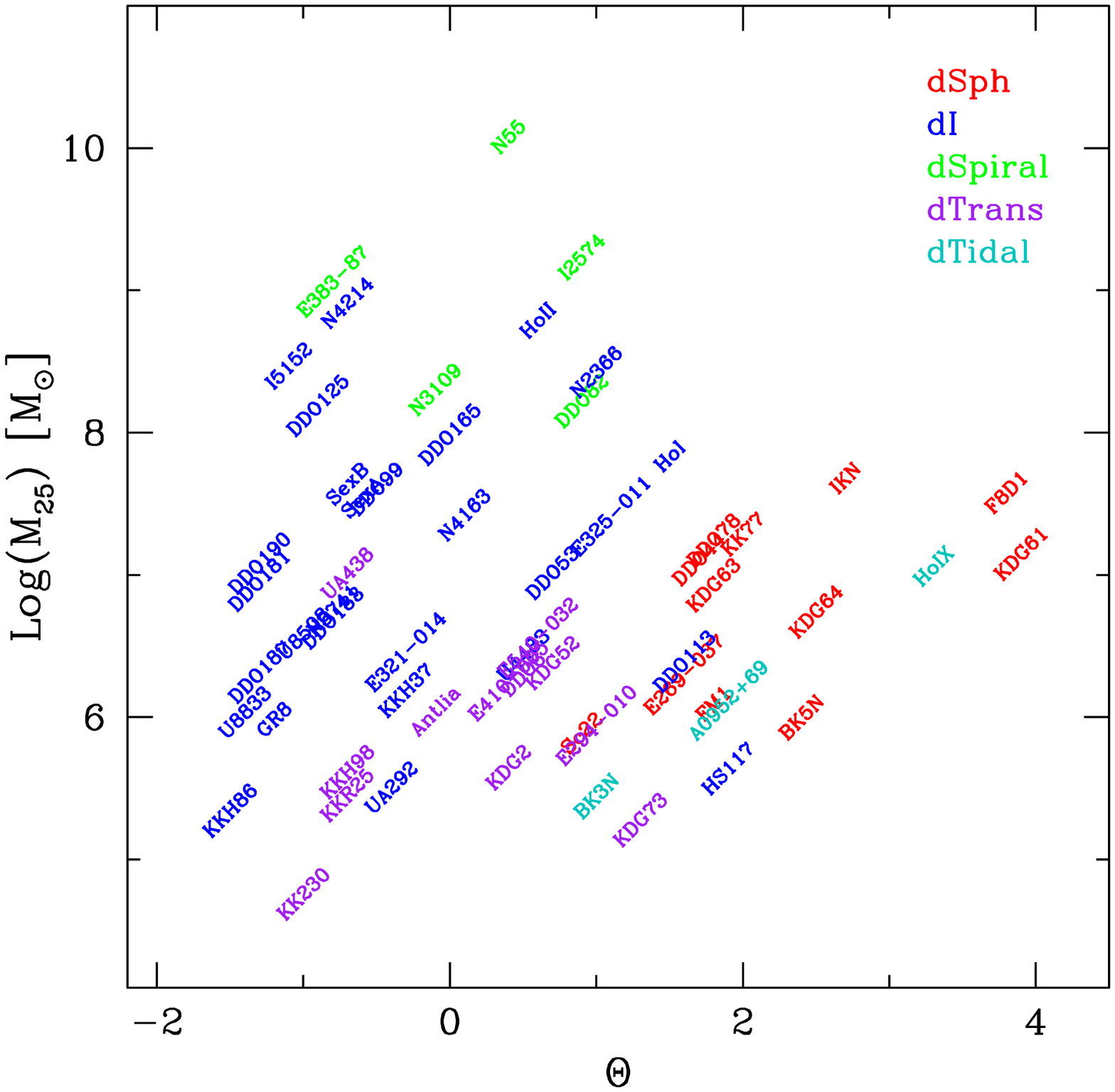}
\caption{The observed morphology--density relationship for the ANGST dwarf galaxies. Values of $M_{25}$ have been derived by integrating the SFHs in Figure \ref{sfh1}, and normalizing to $\mathcal{A}_{25}$ to account for differences in the observed areas.  The tidal indices, $\Theta$, have been taken from \citet{kar04}.  Negative values of $\Theta$ represent isolated galaxies, while positive values represent typical group members. } 
\label{morphden}
\end{center}
\end{figure*}

\section{The Morphology-Density Relationship}
\label{morph}

Galaxies in clusters, groups, and the field follow similar morphology--density relationships.  Namely, gas-poor galaxies, i.e., ellipticals, are generally found to be less isolated than gas-rich galaxies, i.e., spirals \citep[e.g.,][]{dre80, oem74, bla05}.

\begin{figure*}[t]
\begin{center}
\epsscale{0.9}
\plotone{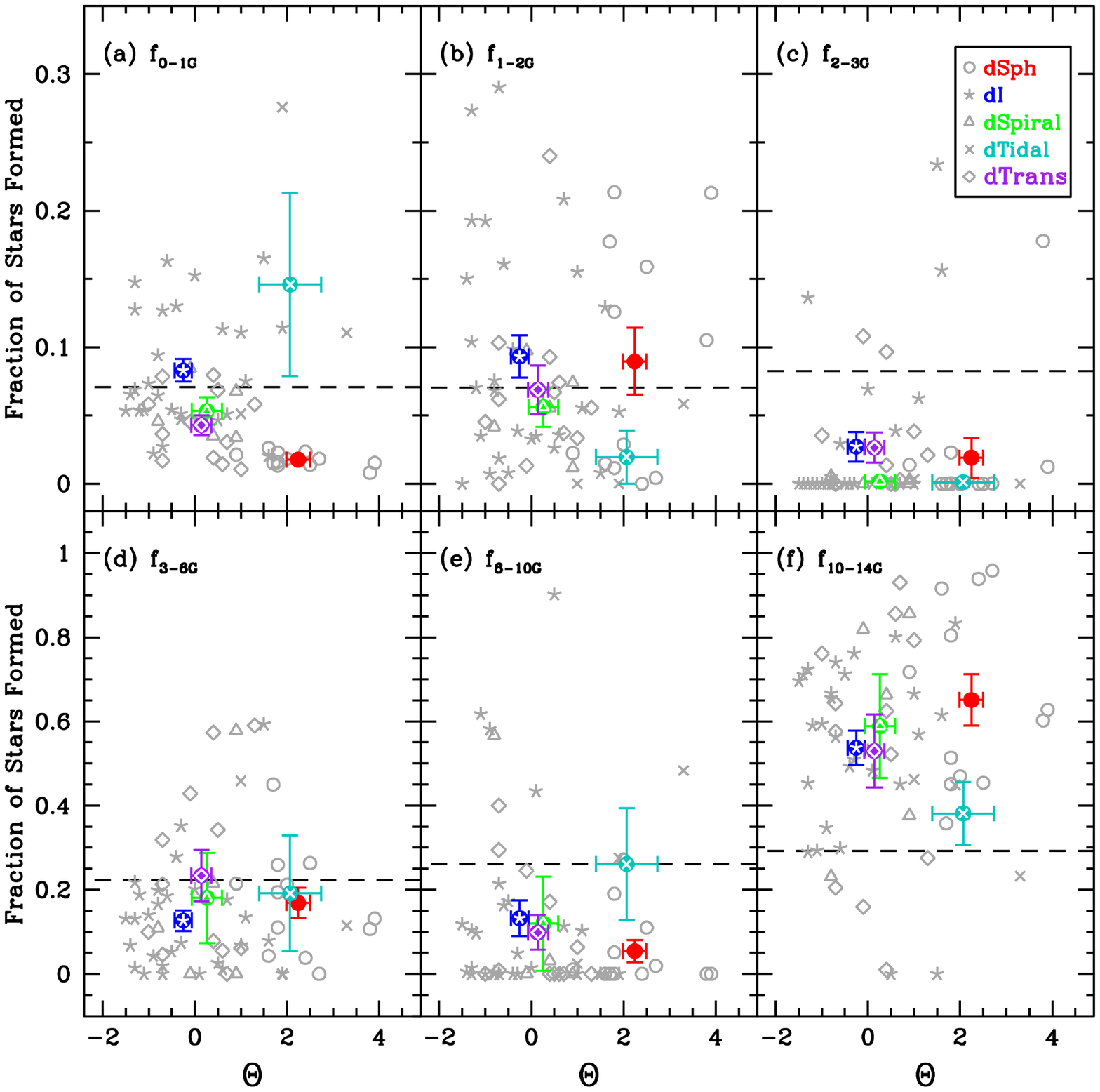}
\caption{The dwarf galaxy morphology--density relationship represented by the fractional SFH, i.e., the fraction of total stellar mass formed during a given time bin, as a function of tidal index, $\Theta$, and morphological type. The time bins are such that $f_{0-1G}$ $=$ 0-1, $f_{1-2G}$ $=$ 1-2, $f_{2-3G}$ $=$ 2-3, $f_{3-6G}$ $=$ 3-6, $f_{6-10G}$ $=$ 6-10, $f_{10-14G}$ $=$ 10-14 Gyr ago. The tidal indices, $\Theta$, have been taken from \citet{kar04}. Negative values of $\Theta$ represent isolated galaxies, while positive values represent typical group members.  The grey points represent individual galaxies, while the larger colored points are the mean values.  Error bars are the uncertainties in the mean values.}  
\label{thf_avg}
\end{center}
\end{figure*}

This same morphology--density relationship has been found for dwarf galaxies \citep[e.g.,][]{ein74, mat98, van00, ski03a, geh06}.  The change in the relative fraction of gas-rich and gas-poor galaxies with environment, provides a simple test for various models of dwarf galaxy evolution. As illustrated in Figures \ref{morphden} and \ref{thf_avg}, the ANGST dwarf galaxies clearly adhere to the morphology--density relationship, when using tidal index, $\Theta$, as a proxy for local density. Typically, dIs are significantly more isolated than dSphs, in spite of having similar mean total stellar masses.  dTrans, on average, have intermediate tidal indices between dIs and dSphs (see \S \ref{transition}), and have a lower mean stellar mass than either. dSpirals are the most massive galaxies, and are typically located in regions of intermediate isolation.   These findings are in general agreement with earlier studies of LG dwarf galaxies \citep[e.g.,][]{mat98, van00, tol09}.

When combined with measured SFHs, the morphology--density relationship can be used to gauge the reliability of dwarf galaxy evolution models, particularly processes that induce gas loss.  Scenarios favoring internal mechanisms (i.e., stellar feedback) as the primary driver of gas loss \citep[e.g.,][]{dek86, dek03} can reproduce a number of observed dwarf galaxy properties \citep[e.g., surface brightness, rotation velocities, metallicities, etc;][]{woo08}.  However, such models are generally unable to account for the morphology--density relationship \citep[e.g.,][]{may09} and often predict that gas-rich and gas-poor dwarf galaxies may have different patterns or efficiencies of SF \citep[e.g.,][]{dek86, ski95}.  In contrast, some models that additionally factor in external effects (e.g., gravitational interactions, ram pressure stripping) have been able to reproduce a wide range of dwarf galaxy properties, including a canonical morphology--density relationship \citep[e.g.,][]{may06, may09, kaz10}.  In the following sections, we discuss results from the ANGST SFH analysis within the context of physical processes that can affect the evolution of dwarf galaxies.

\subsection{Comparing Stellar and Gas Masses}
\label{gascomp}

The relative masses of the gas and stellar components can provide clues to dwarf galaxy evolution. For the ANGST sample, we consider the gas mass and baryonic gas fraction as functions of total stellar mass and tidal index.  The gas masses are based on the \hi\ masses in \citet{kar04}, corrected by a factor of 1.4 to account for helium content, while the baryonic gas fractions, $M_{gas}$/$M_{baryonic}$, is defined to be $M_{gas}$/ ($M_{gas}$+$M_{25}$), and does not account for the contribution due to warm/hot baryons.  The stellar masses, $M_{25}$, are derived from integrating the SFHs over time and applying the areal normalization, $\mathcal{A}_{25}$.

\newpage

\begin{figure*}[t]
\begin{center}
\epsscale{0.9}
\plotone{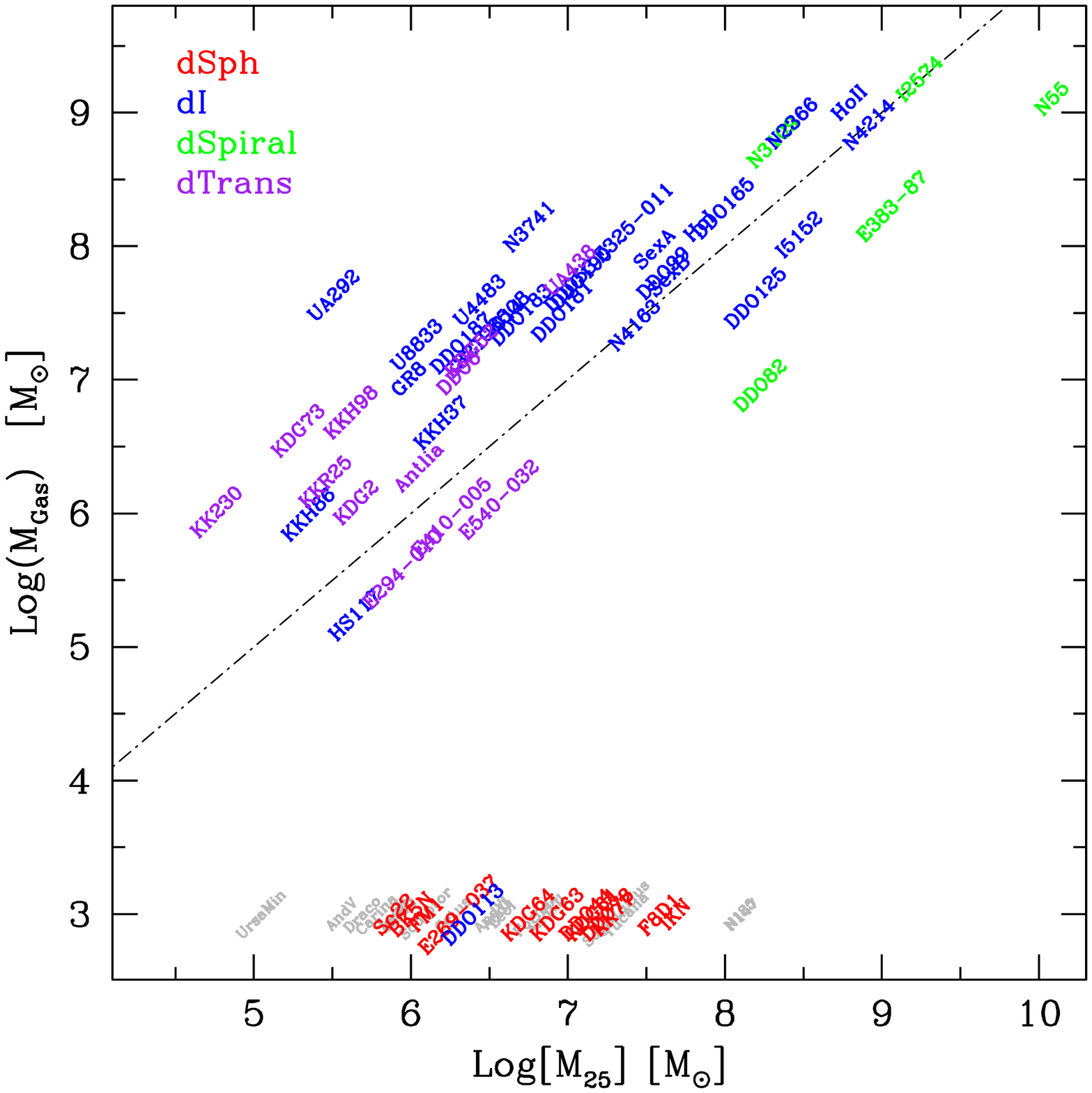}
\caption{\small{Log($M_{Gas}$) plotted versus Log($M_{25}$), where $M_{25}$ has been derived by integrating the SFHs (Figure \ref{sfh1}), and normalizing to $\mathcal{A}_{25}$ to account for differences in the observed areas. We computed $M_{Gas}$ by correcting the \hi\ masses from \citet{kar04} by a factor of 1.4 to account for helium content. The dot-dashed line represents the $M_{Gas}$ $=$ $M_{25}$ equality.  Galaxies without detectable gas are located in the lower portion of the plot, and have been placed at $\log(M_{gas}) =$ 3 for convenience. To illustrate the range of stellar masses spanned by the ANGST sample, we have included select LG dSphs in grey. Considering a present day gas-rich galaxy, SF will increase the stellar mass and decrease the gas mass, moving the galaxy down and to the right. Gas removal will move a galaxy downward, while gas addition, e.g., accretion, moves a galaxy up.  Stellar mass loss moves a galaxy to the left.  SF and stellar feedback alone cannot transform a gas-rich galaxy into a dSph, however models including external effects, e.g., ram pressure stripping and tidal forces, can remove sufficient stellar and gas mass to transform a gas-rich into a gas-poor galaxy (see \S \ref{gasloss}).}}
\label{masses}
\end{center}
\end{figure*}

We first consider the relationship between total gas mass and integrated stellar mass as shown in Figure \ref{masses}, where the dot-dashed line denotes $M_{gas}$ $=$ $M_{25}$. In this context, the most massive galaxies occupy the upper right portion of the plot, while lower mass dwarfs are located to the left.  Galaxies without detectable gas are located in the lower portion of the plot, and have been placed at $\log(M_{gas}) =$ 3 for convenience. To illustrate the range of stellar mass covered by the ANGST dSphs relative to the LG dSphs, we have included select LG dSphs in grey (placed using stellar mass-to-light values from \citealt{mat98}).

Figure \ref{masses} provides a concise snapshot of the current evolutionary state of nearby dwarf galaxies.       For large stellar masses, this view illustrates the morphological ambiguity between dIs and dSpirals;  galaxies from both classes are gas-rich and can have stellar masses $\gtrsim$ 10$^{8}$ \msun.  \hi\ surface density maps of some dIs even reveal hints of spiral gas structure in dIs \citep[e.g.,][]{puc92}.  In this same mass regime, however, there is a  conspicuous absence of dSphs.  In the ANGST sample, we not do find dSphs more massive than $\sim$ 10$^{8}$ \msun, which is in agreement with the lack of massive gas-poor galaxies in the LG \citep[e.g.,][]{mat98, tol09}.  dTrans are predominantly located at low stellar masses, yet have relatively large gas supplies.  These properties suggest that some dTrans may not be significantly different from low mass dIs; we return to this point in \S \ref{transition}.  

\begin{figure*}[t]
\begin{center}
\epsscale{0.9}
\plotone{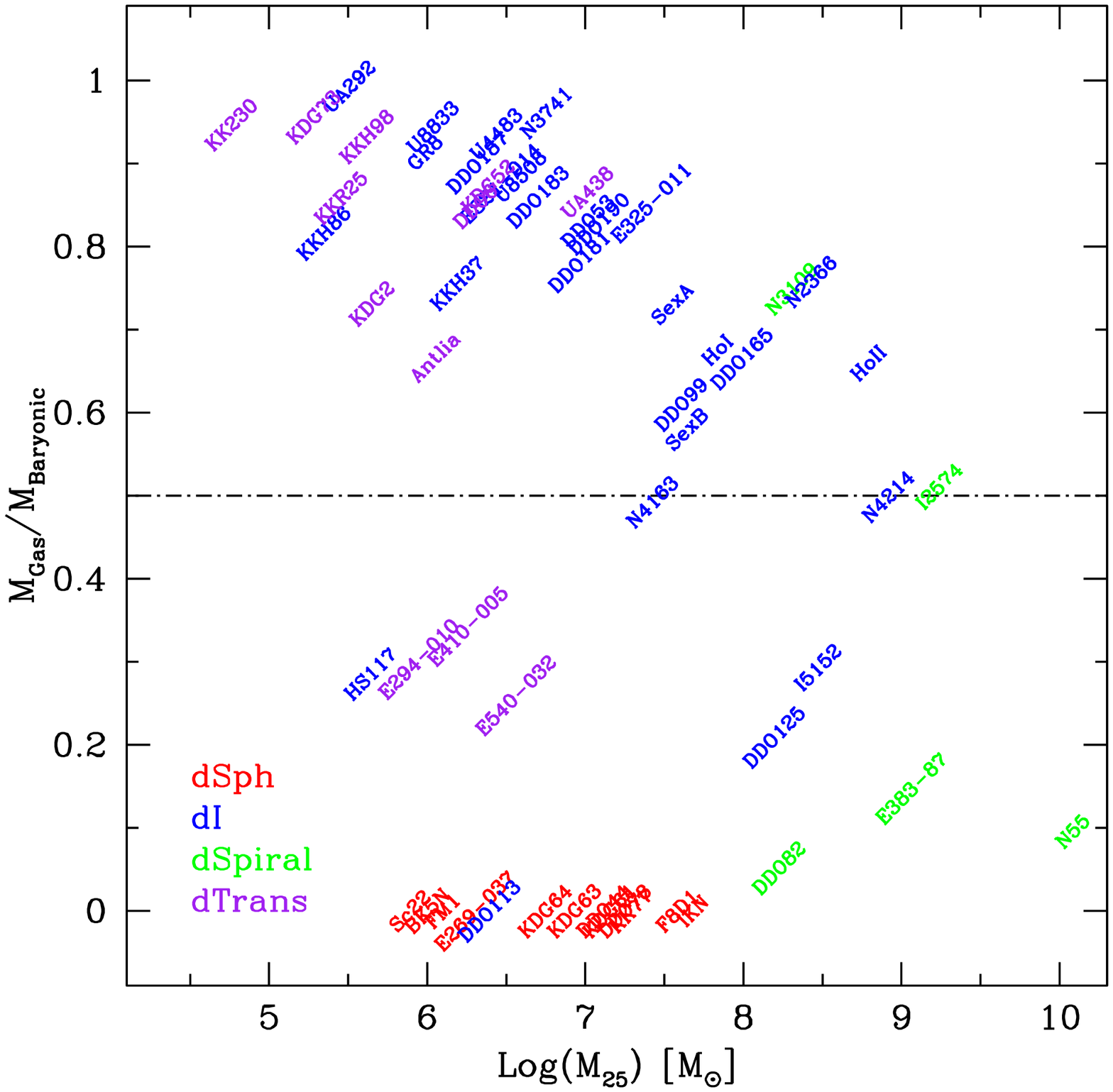}
\caption{The ratio of gas mass to total baryonic mass (i.e., the gas fraction; $M_{Gas}$ to $M_{Gas}$ $+$ $M_{25}$) plotted versus Log($M_{25}$).  We computed $M_{Gas}$ by correcting the \hi\ masses from \citet{kar04} by a factor of 1.4 to account for helium content. For convenience, gas-poor galaxies have been placed at Log($M_{gas}$) $=$ 3. The total integrated stellar masses are from the SFHs (Figure \ref{sfh1}), normalized to $\mathcal{A}_{25}$ to account for differences in the observed areas. The dot-dashed line represents the $M_{Gas}$ $=$ $M_{25}$ equality.}
\label{bary}
\end{center}
\end{figure*}

The baryonic gas fraction provides a more detailed view of the current evolutionary state of the ANGST dwarf galaxies (Figure \ref{bary}).   The general trend for dwarf galaxies mirrors that of massive counterparts on the Hubble sequence \citep[e.g.,][]{rob94}, namely that spiral galaxies typically have low gas fractions, while irregulars have higher gas fractions.  This perspective also reinforces the view that many dTrans could be low mass dIs in between episodes of massive SF \citep[e.g.,][]{ski03a}. 

There appear to be several outliers to the general trends in Figure \ref{bary}. The two most conspicuous outliers, HS~117 and DDO~113, may simply be morphological misclassifications. HS~117 is classified as a dI, but has a very low gas content, with an upper limit of $M_{\hi}$ $\sim$ 10$^{5}$ \msun\ \citep{huc00} and error bars consistent with zero.  Interestingly, \citet{kar07} detect low levels of  \halpha\ in HS~117, which reinforces the dI classification, but based on the lack of \hi, they deem this a dSph.  Morphologically, inspection of the HST image further reveals that it appears to be superimposed on an \hii\ region, leading to an erroneous dI classification.  We therefore suggest HS~117 is best classified as a dSph or possibly a dTrans.  DDO~113 is likewise classified as a dI with negligible gas content, but has a possible \halpha\ detection \citep{kar07}.   Inspection of the HST based CMD suggests that DDO~113 resembles a prototypical dSph, with a handful of relatively faint blue stars, suggesting it is also likely a dSph or a dTrans.

Additional putative outliers include three dTrans (ESO294-010, ESO410-005, ESO540-032) and three dIs (NGC~4163, IC~5152 and DDO~125).  Two of the dIs  (IC~5152 and DDO~125) are isolated and relative massive, yet have moderate gas fractions.  The three dTrans have confirmed blue horizontal branch populations \citep{daC10}, which helps constrain their ancient SFHs. However, none of these galaxies show any unique features in their SFHs that would explain their outlier status.  In contrast, NGC~4163 is a starburst dwarf galaxy, which may have consumed a significant amount of gas due to an intense burst of SF within the last $\sim$ 1 Gyr \citep[e.g.,][]{mcq09}.  

Examining the gas fraction as a function of isolation (Figure \ref{thbary}), we see that isolated galaxies tend to have high gas fractions, while those in high density environments have low gas fractions.  Interestingly, there appear to be no galaxies with \emph{any} appreciable gas in dense environments ($\Theta$ $\gtrsim$ 1.5; HS~117 and DDO~113 have uncertain gas measurements as described above).  We also see some evidence for separation of dTrans into groups of isolated ($\Theta$ $<$ 0) and moderate/high ($\Theta$ $\lesssim$ 0) density environments, which we further discuss in \S \ref{transition}.

\subsection{Mechanisms for Complete Gas Loss}
\label{gasloss}

How gas-rich galaxies \emph{completely} lose their gas has long been an outstanding question in dwarf galaxy evolution \citep[e.g.,][and references therein]{may09}.  Within the ANGST sample, we can test the viability of various gas loss mechanisms using the observed properties of gas-poor dSphs.  The ANGST dSphs share several notable characteristics including: (1) little SF in the most recent 1 Gyr compared to gas-rich dwarfs, (2) similar total stellar masses, (3) SFHs that are generally extended and indistinguishable from dIs, and (4) are located exclusively in high density environments.  In what follows, we consider the impact of putative gas loss mechanisms on the evolution of a typical gas-rich dwarf galaxy (e.g., with both $M_{25}$ and $M_{gas}$ $\sim$ 10$^{7}$ \msun) in the context of Figure \ref{masses}.

The first mechanism for gas removal is consumption through SF. This process will increase the stellar mass and decrease the gas mass of our prototypical dI, moving it to the right on Figure \ref{masses}.  Though SF can consume large amounts of a galaxy's gas reservoir, the gas densities will eventually become too low to continue to form stars \citep[e.g.,][]{ken89}.  This suggests that if consumption was the only mechanism, we should observe trace amount of gas in dSphs, which is not typically the case.

In the event that gas density was not a limiting factor, characteristically low SFRs and SF efficiencies in dwarf galaxies \citep[$\sim$ 1\%; e.g.,][]{ler08} imply long gas consumption timescales.  In this event, gas supplies would typically not be exhausted for more than a Hubble time, which suggests that dSphs would be quite rare. In addition, gas consumption does not have a dependence on the environment external to a galaxy. Thus, it cannot account for the observed morphology--density relationship. Although SF is a critical process to galaxy evolution, it cannot be responsible for the complete removal of gas from dwarf galaxies.

The second possible mechanism for removing gas is stellar feedback.  Mechanical energy due to stellar winds and supernovae provide an appealing explanation for gas removal from a galaxy via expulsion \citep[e.g.,][]{dek86}.  For the prototypical gas-rich dI, this process removes gas with minimal effect on stellar mass, moving the galaxy straight down in Figure \ref{masses}.  Several feedback models can reproduce observed dwarf galaxy relationships between mass, luminosity, and rotational velocities \citep[e.g.,][]{dek86, dek03, woo08}.  Additionally, these models are generally able explain the observed mass--metallicity relationship for dwarf galaxies \citep[e.g.,][]{lee06}.  

Although these models hold promise, there are two significant challenges to gas removal in dwarf galaxies due to feedback.  First, a number of simulations have demonstrated that the energy due to stellar feedback is insufficient to completely expel cold gas (e.g., \hi) from the gravitational potential of a typical dwarf galaxy \citep[e.g.,][]{mac99, der99, mar06, rev09}.  Second, stellar feedback alone cannot explain the morphology--density relationship.  Similar to the rationale for gas consumption through SF, stellar feedback is not dependent on the environment surrounding a galaxy.  Thus, while SF and stellar feedback are primary drivers of dwarf galaxy evolution, it is unlikely that either or both are able to account for the final transition from a gas-rich to a gas-poor state.

\begin{figure*}[t]
\begin{center}
\epsscale{0.9}
\plotone{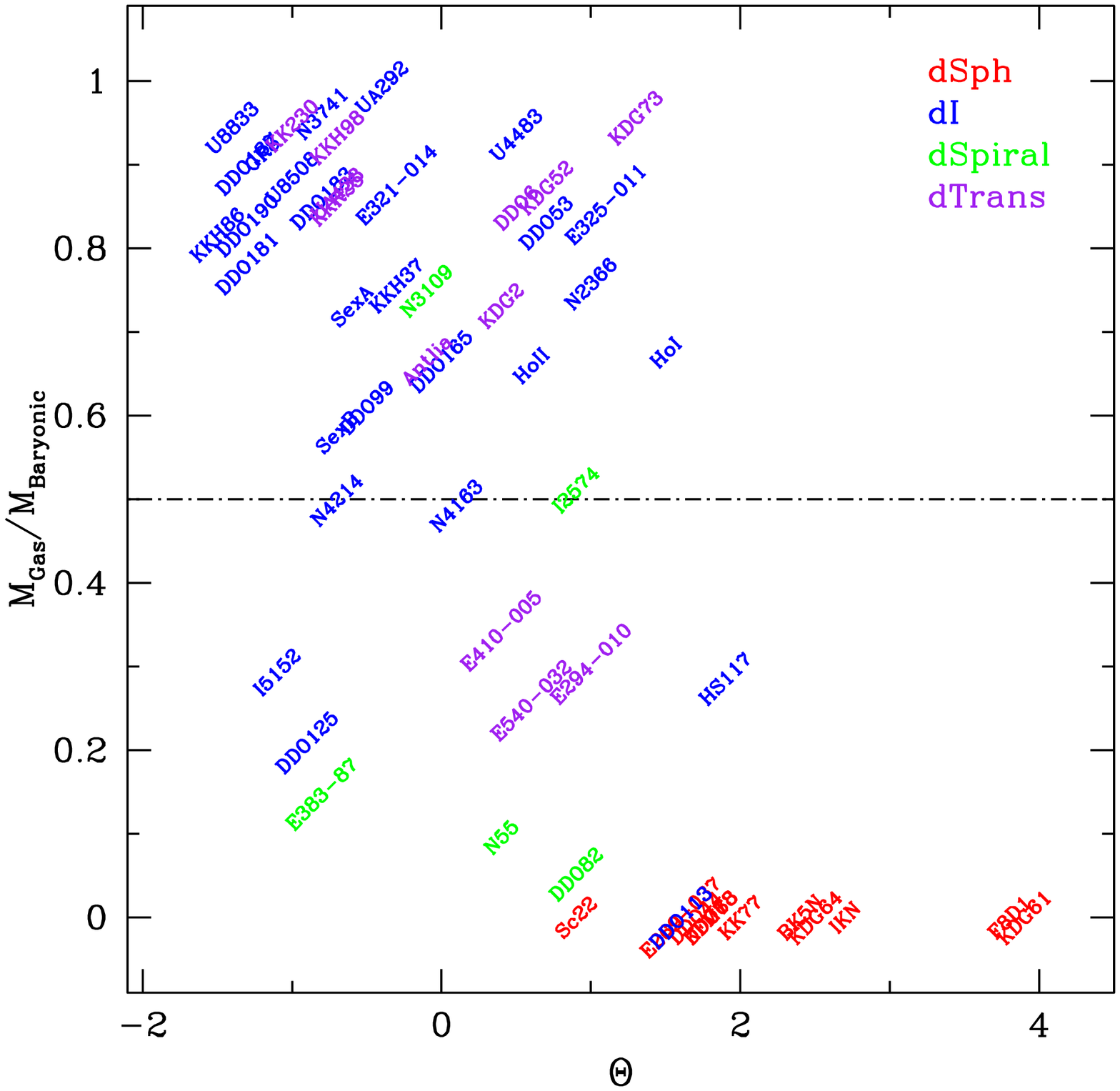}
\caption{The ratio of gas mass to total baryonic mass (i.e., the gas fraction; $M_{Gas}$ to $M_{Gas}$ $+$ $M_{25}$) plotted versus the tidal index, $\Theta$.  The tidal indices, $\Theta$, have been taken from \citet{kar04}. Negative values of $\Theta$ represent isolated galaxies, while positive values represent typical group members.  \hi\ masses have been taken from \citet{kar04}. We computed $M_{Gas}$ by correcting the \hi\ masses from \citet{kar04} by a factor of 1.4 to account for helium content. The total integrated stellar masses are from the SFHs (Figure \ref{sfh1}), normalized to $\mathcal{A}_{25}$ to account for differences in the observed areas.  Gas-poor galaxies typically have positive tidal indices, while gas-rich galaxies have predominantly negative values.  It is interesting to note that for $\Theta$ $\gtrsim$ 1.5, there are no gas-rich galaxies.} 
\label{thbary}
\end{center}
\end{figure*}

The third mechanism we consider is ram pressure stripping.  \citet{lin83} first proposed that dSphs were once gas-rich dwarfs whose gas supply has been removed by ram pressure stripping from the intergalactic medium (IGM).  This mechanism has negligible impact on stellar mass loss, thus moving our prototypical dI straight down in Figure \ref{masses}.  Satellite galaxies in higher density environments  would likely encounter a denser IGM, which would lead to more efficient gas loss.  This provides a feasible explanation for the observed morphology-density relationship.  

Despite the appeal of this mechanism, there is a distinct lack of observational evidence of systematic ram pressure stripping of LG dwarf galaxies.  For example, \citet{mcc07} and \citet{kni09} find that ram pressure stripping appears to be a localized phenomena, and only evident for a small minority of LG dwarfs.  However, given that the LG is in a state of relatively passive evolution, it could be that signatures of ram pressure stripping have been erased in many satellites.  An additional challenge comes from the magnitude of ram pressure stripping. \citet{may06} demonstrate the ram pressure stripping is unlikely effective enough to completely remove the gas supply of a dwarf galaxy in the LG.  It appears that ram pressure stripping is not able to account for complete gas loss in dwarf galaxies.

We next consider tidal effects on gas removal from dwarf galaxies.  The close passage of a dwarf galaxy to a massive companion has strong gravitational effects on the gas, stellar, and dark matter contents of the smaller galaxy.  As shown in \citet{may01b}, the magnitude of the tidal force during an interaction does not appear to be enough to remove the gas content from a gas-rich dI.  Instead, \citet{may01a, may06} advocate a more complex approach in which a combination of ram pressure stripping and stellar mass loss due to tidal effects provide a plausible model, `tidal stirring', for the transformation of gas-rich dIs into gas-poor dSphs.  In this scenario, the prototypical gas-rich dI is able to lose both stellar and gas mass, allowing it to move down and left in Figure \ref{masses}, meaning a dI with a high stellar mass could be transformed into a less massive dSph.

Predictions from tidal stirring models provide qualitative explanations for a number of trends seen in the ANGST sample.  Foremost, tidal stirring naturally produces a morphology--density relationship.  Galaxies in lower density environments have had few (or no) interactions with massive galaxies, and are able to maintain high gas fractions.  This is in general agreement with the trends seen in Figure \ref{thbary}.  Of the galaxies that do interact with a massive companion, tidal stirring predicts that stellar and gas mass loss happens progressively over several Gyr \citep[a single complete orbit in the LG is typically 1-2 Gyr;][]{kaz10}.  Therefore, these galaxies spend the majority of their life cycling between SF and stellar feedback.   By extension, this implies that many dSphs and dIs should not have drastically different SFHs, in general agreement with the SFHs of ANGST dwarf galaxies.   

While the general qualitative agreement between tidal stirring predictions and the ANGST sample is encouraging, a quantitative comparison between predictions and observed dwarf galaxy SFHs are needed to place precise constraints on the gas loss processes in dwarf galaxies.

\subsection{The Nature of Transition Dwarf Galaxies}
\label{transition}

With little evidence of recent SF, yet detectable amounts of \hi, dTrans may hold clues to the transformation of gas-rich to gas-poor galaxies. Synthesizing several previous studies, there are two favored scenarios for the origins of dTrans, namely that they are either in the last throes of SF or are observed during temporary lulls between episodes of massive star formation \citep[e.g.,][]{cot97,mat98, gre03, ski03a, cot09}. 

dTrans in the ANGST sample have characteristics consistent with both scenarios.  Those with low gas fractions ($M_{gas}$/$M_{baryonic}$ $\lesssim$ 0.4; ESO294-010, ES410-005, and ESO540-032) typically have positive tidal indices, indicating a higher likelihood of gas loss via interactions (Figure \ref{thbary}).  These three galaxies may be examples of galaxies genuinely transitioning from gas-rich to gas-poor states.  In contrast, the most isolated dTrans (KK~230, KKH~98, KKR~25, and UA~438) typically have high gas fractions ($\gtrsim$ 0.9) and stellar masses among the lowest in the ANGST sample (Figure \ref{bary}).  These dTrans have properties consistent with other low mass dIs, and could simply be low mass dIs lacking \halpha, which can be attributed to a wide range of effects including temporary episodes between massive star formation \citep[e.g.,][]{ski03a}, a stochastically sampled or systematically varying IMF \citep[e.g.,][]{eld10, wei05}, and/or leakage of ionizing photons \citep[e.g.,][]{fer96}.  The remaining dTrans (Antlia, KDG~2, DDO~6, KDG~52, KDG~73) appear to be have masses and tidal indices close to the genuinely transitioning group, but gas fractions similar to the low mass dIs.  It is possible that these galaxies may have had few interactions, resulting in less gas loss.  Detailed multi-wavelength studies of these galaxies could provide insight into gas loss mechanisms in dwarf galaxies.

\section{Dwarf Galaxies in a Cosmological Context}
\label{cosmic}

As the smallest and most pristine galaxies in the Universe, dwarf galaxies occupy a critical, but poorly constrained role in cosmological models of galaxy formation and evolution \citep[e.g.,][]{moo99, kly99}. Because of their intrinsic faintness, dwarf galaxies have been difficult to directly detect at cosmologically significant redshifts \citep[e.g.,][]{mat04}.  

\begin{figure}[t]
\begin{center}
\epsscale{1.1}
\plotone{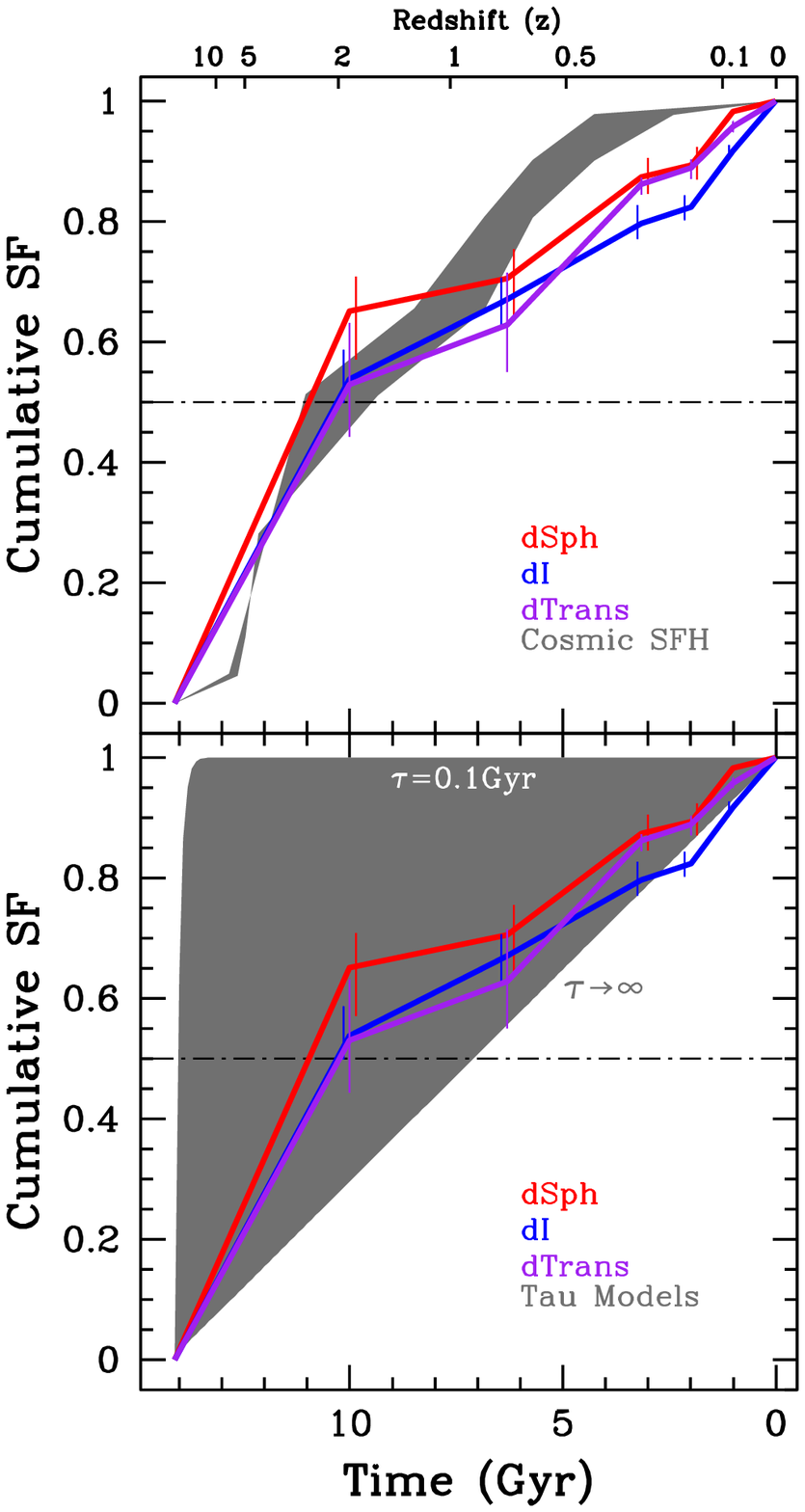}
\caption{\small{The mean cumulative SFHs for dSphs (red), dIs (blue), and dTrans (purple).  The error bars represent the uncertainties in the mean , while the dot-dashed line indicates 50\% of the total stellar mass.  In the top panel, the grey shaded region represents the cosmic SFH as measured by \citet{red08}.  We see general good agreement between the lifetime SFHs of dwarf galaxies and the cosmic SFH, with some discrepancies at intermediate times (see \S \ref{cosmic} for further discussion).  In the bottom panel, the grey shaded region represents the expected cumulative SFH for an exponentially declining SFH model, for a range of $\tau$ values from 0.1 to 14.1 Gyr.  Notice that single valued $\tau$ models are not good representations of the observed SFHs.}}
\label{cum2}
\end{center}
\end{figure}

As an alternate approach to observing high redshift dwarf galaxies, we present a comparison between the mean cumulative SFHs of the ANGST sample dwarfs and the cosmic SFH, as derived from observed UV fluxes in high redshift galaxies \citep[e.g.,][]{red08}.  In the top panel of Figure \ref{cum2}, we see good agreement between the cosmic SFH (grey shaded region) and the mean cumulative SFHs of the ANGST dwarf galaxies at z $\sim$ 2 (10 Gyr ago).  At more recent times, the cosmic SFH increases sharply, while the cumulative SFHs of dwarf galaxies increase at a slower rate.  

This comparison suggests that the largest differences between SFRs in dwarf galaxies and massive galaxies studied at higher redshifts are at intermediate and recent epochs.  This finding is generally consistent with the expectations of galaxy `downsizing', where high mass galaxies preferentially stop forming stars at higher redshifts \citep[e.g.,][]{cow96}.  The precise role of dwarf galaxies in theories of downsizing is highly uncertain, making this result difficult to interpret in the context of current models \citep[e.g.,][]{mou06}. We also caution that the physical significance of this finding is still ambiguous due to a several factors including uncertain AGB star models and redshift dependent dust corrections applied to the cosmic SFH \citep[e.g.,][]{gir10, mar10, red10}.  A comparison of SFHs from the entire ANGST sample (including the most massive galaxies) and the cosmic SFH is presented in \citet{wil11}.

The cumulative SFHs can also provide insight into the validity of model SFHs often used to describe the evolution of dwarf galaxies.  As an illustrative example, we have selected a simple exponentially declining SF model (SFR $\propto$ $e^{-t/\tau}$; a $\tau$ model), and show the cumulative SFHs for a range of $\tau$ values (grey shading) in the lower panel of Figure \ref{cum2}.  Visual inspection suggests that the measured SFHs show significant deviations from the predicted smooth curve of single $\tau$ valued models.  A $\chi$$^{2}$ test between the measured cumulative SFHs and exponentially declining SFH models (with $\tau$ varying from 0.1 to 14.1 Gyr) confirms that any single value of $\tau$ does not accurately represent the data.  Similarly, the mean measured SFHs also are not well matched with a constant SFH (i.e., $\tau$ $\rightarrow$ $\infty$ Gyr) or a single ancient epoch of SF followed ($\tau$ $\lesssim$ 0.1 Gyr) by passive galaxy evolution.   Instead more complex and multi-component models may better describe the lifetime SFHs of dwarf galaxies.

\section{Summary}

We uniformly analyzed SFHs of 60 dwarf galaxies in the nearby universe based on observations and data processing done as part of the ANGST program.  While the SFHs of individual galaxies are quite diverse, we find that the mean SFHs of the different morphological types are generally indistinguishable outside the most recent $\sim$ 1 Gyr.  On average, the typical dwarf galaxy formed $\sim$ 50\% of its stellar mass by z $\sim$ 2 and 60\% by z $\sim$ 1.  

Among the morphological types, the SFHs hint at divergence within the past few Gyr, although assigning a precise time to this phenomena is challenging due to uncertainties in AGB star modeling and the modest time resolution afforded by the data.  The clearest differences between the morphological types can been seen in the most recent 1 Gyr, where the typical dSph, dI, dTrans and dSpiral formed $\sim$ 2\%, 8\%, 4\%, and 5\% of their total stellar mass.  

The dwarf galaxies in the ANGST sample show a strong morphology--density relationship.  This suggests that internal mechanisms, e.g., stellar feedback, cannot solely account for gas-loss in dwarf galaxies, as the corresponding models are unable to produce this observed relationship. Instead, we find qualitative consistency with the model of `tidal stirring' \citep[e.g.,][]{may01a, may01b, may06}, which can broadly explain the extended SFHs as well as the observed morphology--density relationship.  

A comparison with the cosmic SFH reveals general good agreement between the independently derived SFHs.  The slower rate of SF for dwarf galaxies at intermediate epochs may be tentative evidence of galaxy downsizing.  However, broad uncertainties in extinction corrections and AGB star models may be able to explain the offset. Further, the mean measured SFHs are inconsistent with single valued exponential models of SF (i.e., $\tau$ models), and may require more complex or multi-component models.

We also identify 12 dTrans in the sample, based on the literature definition of present day gas fraction and SF as measured by \halpha\ \citep[e.g.,][]{mat98}.  Within this sample of dTrans, we find that galaxies with high gas fractions are associated with more isolated galaxies, while those with lower gas fractions are less isolated.  This suggests that there are two mechanisms that can produce the observed dTrans characteristics: the isolated gas-rich galaxies are simply between episodes of SF due to the stochastic nature of SF in low mass galaxies, while the less isolated galaxies could be in the process of interacting with a more massive companion. 
 
\section{Acknowledgments}

DRW is grateful for support from the University of Minnesota Doctoral Dissertation Fellowship and Penrose Fellowship. IDK is partially supported by RFBR grant 10-02-00123. The authors would like to thank Stephanie C{\^o}t{\'e} for fruitful discussions on the nature of transition dwarf galaxies and Oleg Gnedin for his insightful suggestions on measures of photometric quality. This work is based on observations made with the NASA/ESA Hubble Space Telescope, obtained from the data archive at the Space Telescope Science Institute.  Support for this work was provided by NASA through grants GO-10915, DD-11307, and GO-11986 from the Space Telescope Science Institute, which is operated by AURA, Inc., under NASA contract NAS5-26555. This research has made use of the NASA/IPAC Extragalactic Database (NED), which is operated by the Jet Propulsion Laboratory, California Institute of Technology, under contract with the National Aeronautics and Space Administration.  This research has made extensive use of NASA's Astrophysics Data System Bibliographic Services.  

{\it Facility:} \facility{HST (ACS, WFPC2)}

\clearpage

\appendix

\section{Optimizing Time Resolution of CMD-based SFHs} 
\label{appena}

Meaningful comparison of SFHs across the ANGST sample requires a uniform time binning scheme.  This necessitates balancing the wealth of information about the ancient and intermediate epochs from the deepest CMDs (e.g., Antlia; 50\% completeness of $M_{F814W}$ $\sim$ $+$1.75) with the course leverage of the shallower CMDs (e.g., UGC~4483; 50\% completeness of $M_{F814W}$ $\sim$ $-$1.7; see Table \ref{tab2}).  As a compromise between the two extremes, we fixed a sample-wide time binning scheme using a galaxy with a 50\% completeness value near the median of the sample (DDO~6; $M_{F814W}$ $\sim$ 0). To test for the optimal time resolution, we constructed CMDs of single age stellar populations, convolved the simulated photometry with the artificial stars and photometric limits of DDO~6, and ran the SFH recovery program on each of these simulated CMDs. This process was conducted on single age CMDs of 0.5, 1.5, 2.5, 4.5, 8, and 12 Gyr in age.

\begin{figure*}[t]
\begin{center}
\epsscale{1.0}
\plotone{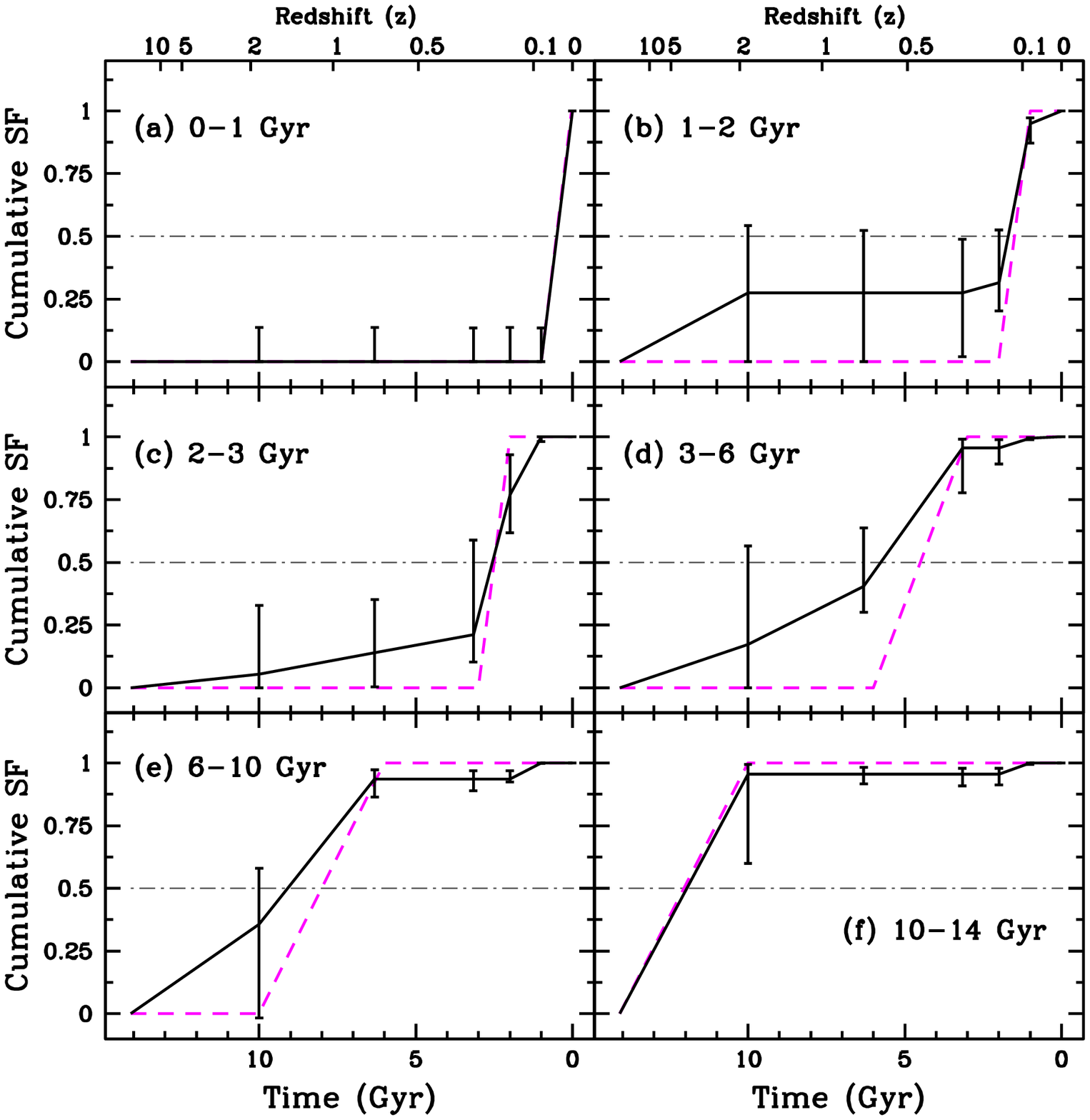}
\caption{\small{A comparison between simulated input single age cumulative SFHs and the recovered SFHs (solid black line), demonstrating the time binning scheme for the ANGST sample.  Here the magenta dashed line indicates the shape of a perfectly recovered cumulative SFH in the correct time bin.  Delta function SFHs were created at 0.5, 1.5, 2.5, 4.5, 8, and 12 Gyr ago and convolved with the artificial star tests of DDO~6, a galaxy representative of the median photometric depth in the ANGST sample.  These single age CMDs were then recovered in a manner identical to the SFH recovery method, including 50 Monte Carlo realizations for error analysis.  The grey dot-dashed line shows indicates 50\% of the total stellar mass formed.  We find general consistency between the input and recovered SFHs, particularly in the 0-1 and 10-14 Gyr time bins. Intermediate ages show a lower fraction of stars recovered in the desired bin, however all have recovery fraction $\gtrsim$ 50\%.  This broad time binning scheme allows us to balance a variety of photometric depths across the sample with temporal information about patterns of SF.  See Appendix \ref{appena} for a more detailed discussion.}}
\label{frac_sf}
\end{center}
\end{figure*}

The purpose of this exercise is to provide a simple test for the robustness of the final time bins. While coarser time bins encompass a larger fraction of the input SF, there are generally fewer of them, which provides less information about pattens of SF over time.  In this test, we deemed a good recovery as one in which the input SFH was within the uncertainties of the recovered SFH.  In Figure \ref{frac_sf}, we show the cumulative SFHs, where the input SF is indicated by the dashed magenta line and the recovered SFH by the solid black line.  Uncertainties were computed using 50 Monte Carlo realizations as described in \S \ref{sfhs}, including identical shifts in $M_{bol}$ and $T_{eff}$.

Initial tests suggested that six broad time bins of 0-1, 1-2, 2-3, 3-6, 6-10, 10-14 Gyr would provide suitable time resolution across the ANGST sample.  As shown in Figure \ref{frac_sf}, the general agreement between the input and recovered SFHs is good and falls within the uncertainties.  Clearly, the best recovery is seen in the 0-1 Gyr bin, where the young MS and massive core and blue helium burning stars provide secure leverage on recent SF.  The oldest bin also shows similarly good results.  

The results in the intermediate bins are generally reasonable, although not as reliable as the youngest and oldest bins.  Much of this can be attributed to uncertainties in AGB star models \citep[e.g.,][]{gir10, mar10}.  Although, we used the same models to generate and recover the single age populations, photometric depth, systematic stellar model uncertainties, and blurring of distinct features due to observational effects all contribute to uncertainties in the recovered SFH (see Appendix \ref{systematics}).  AGB populations of different ages often have similar colors and magnitudes on optical CMDs \citep[e.g.,][]{gza05, wei08}, thus it can be difficult to confidently determine the epoch of SF at intermediate ages (1--10 Gyr ago).  This limitation is clearly illustrated in Figure \ref{frac_sf}, which shows broad uncertainties on the recovered SFHs during intermediate times.  However, we still find that the input SFHs are generally within the recovered SFH error bars.  Additionally, $\gtrsim$ 50\% of the SF is recovered in the correct time interval.  Although it would be possible to combine bins at intermediate times to increase the accuracy of the recovered SFRs,  we believe that this scheme will permit future comparisons with SFHs derived using improved AGB star models \citep[e.g.,][]{gir10}.  

For this time binning scheme, we find that reported uncertainties in the SFRs typically decrease for CMDs deeper than DDO~6, and increase for shallower CMDs, in agreement with expectations.  We thus conclude that this scheme is adequate for comparison of SFHs across the ANGST sample.

\section{Exploring Uncertainties in CMD-based SFHs}
\label{systematics}

Photometric depth is an important consideration when interpreting the accuracy of a measured SFH.  Varying photometric depths impact both the number of stars on a CMD and determine the presence of age sensitive CMD features (e.g., ancient MS turn-off, horizontal branch).  Intuitively, a CMD with a brighter photometric limit has less information available than a deeper CMD, and the derived SFH is thus more uncertain.  However, quantifying the precise impact of these uncertainties, such as the amplitude and ages affected, is challenging as varying photometric depths can amplify effects of uncertainties in the stellar models used to measure SFHs. In this section, we explore the effects of photometric depth on the accuracy of SFH recovery.  We demonstrate the effects in two regimes: one where the only variable is photometric depth, and the other where we vary both photometric depth and stellar evolution models.

We first analyze the accuracy of recovered SFHs as a function of \emph{only} photometric depth.  That is, we construct synthetic CMDs at select photometric depths, then attempt to recover the input SFH using identical parameters (e.g., IMF, binary fraction, filter combination) and, in this case, the same stellar evolution models.  More concretely, we constructed CMDs at six different photometric depths ($M_{V} =$ $+$4, $+$2, $+$1, 0, $-$1, $-$2), using a constant SFH ($\log(t)=$7.4-10.15, with a time resolution of $\sim$ 0.1 dex), a fixed metallicity, and the BaSTI stellar evolution models.  Each CMD was populated with $\sim$ 10$^{6}$ stars in order to minimize the contribution of random uncertainties due to the number of stars used to measure the SFR in each time bin; For SFHs measured from observed CMDs, the uncertainties for Poisson sampling of the CMD are already well characterized by our Monte Carlo tests.  We then recovered the SFH of each CMD using the BaSTI stellar evolution models, to ensure that the only variable being tested is photometric depth.

\begin{figure}[hb!]
\begin{center}
\epsscale{1.0}
\plotone{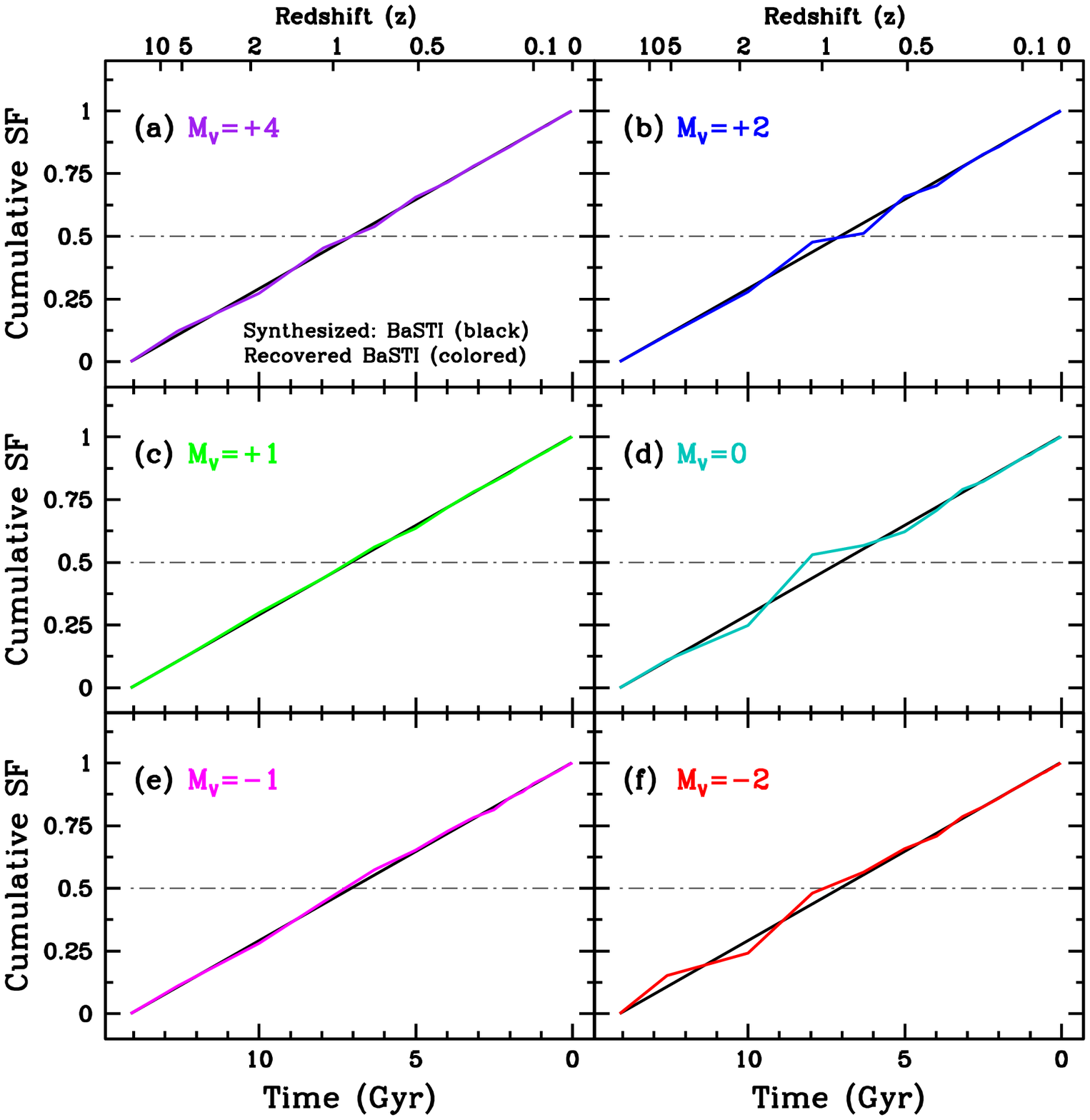}
\caption{\small{A test of the effects of \emph{only} photometric depth on the accuracy of SFH recovery.  The black solid line is the input constant SFH, while the colored lines are the recovered SFHs at six select photometric depths.  Synthetic CMDs were generated using the BaSTI stellar models, a constant SFH, and a fixed metallicity. The SFHs of the synthetic CMDs were then recovered using identical parameters, including the BaSTI stellar models.  The input and recovered SFHs are in excellent agreement at all photometric depths.  This implies that if the underlying stellar model is known exactly, then a SFH can be recovered to a precision limited by the number of stars in the CMD.}}
\label{basbas_sf}
\end{center}
\end{figure}

In Figure \ref{basbas_sf}, we compare the input (black lines) and recovered (colored lines) cumulative SFHs at each photometric depth.  Overall, the recovered cumulative SFHs are in excellent agreement with the input SFHs at all photometric depths.  The maximum deviation between the input and recovered SFHs at any photometric depth is $\sim$ 4\%, which is consistent with the expected Poisson precision of 1/$\sqrt(N)$, where $N$ is the number of stars used to measure the SFR in a given time bin.  This exercise demonstrates that if all the underlying stellar models are known exactly, then the accuracy of the recovered SFH only depends on the number of stars in the CMD, and not the photometric depth.  The same results are found when using different stellar evolution models, e.g., Padova, Dartmouth \citep{dot08}, to conduct this exercise.

Systematic uncertainties are introduced into SFH measurements by uncertainties in the selected stellar models.  The luminosity, color, and number density of specific CMD features provide leverage on the SFR at different epochs.  However, stellar models are not always self-consistent when trying to model multiple observed features \citep[e.g., reproducing colors consistent with observations for both the horizontal branch and RGB; e.g.,][]{gza05}.  The effect on a measured SFH is that the SFR may be systematically shifted into a particular time bin, depending on the stellar model used and photometric depth of the CMD (i.e., which particular age sensitive CMD features are available).  We refer the reader to \citet{gza05} for a comprehensive review on the adequacy of stellar evolution models for reproducing CMD features.

To test for systematic effects on measured SFHs, one ideally wants to test for differences between the stellar models and `truth', that is, stellar population characteristics based on the exact physics governing stellar evolution in nature.  However, current stellar evolution models represent the best physical descriptions we have of nature.  By measuring how different stellar models impact a measured SFH, we can get a sense of the amplitude of systematic uncertainties for SFH recovery. 

To examine the magnitude of systematic effects, we again construct six CMDs at selected photometric depths each containing $\sim$ 10$^{6}$ stars, assuming a constant SFH, fixed metallicity, and the BasTI stellar evolution models.  However, the SFHs are now recovered with the Padova stellar models.  Thus, in this case the differences between the input and recovered SFHs are indicative of the systematic effects introduced by choice in stellar model.

\begin{figure}[b]
\begin{center}
\epsscale{0.8}
\plotone{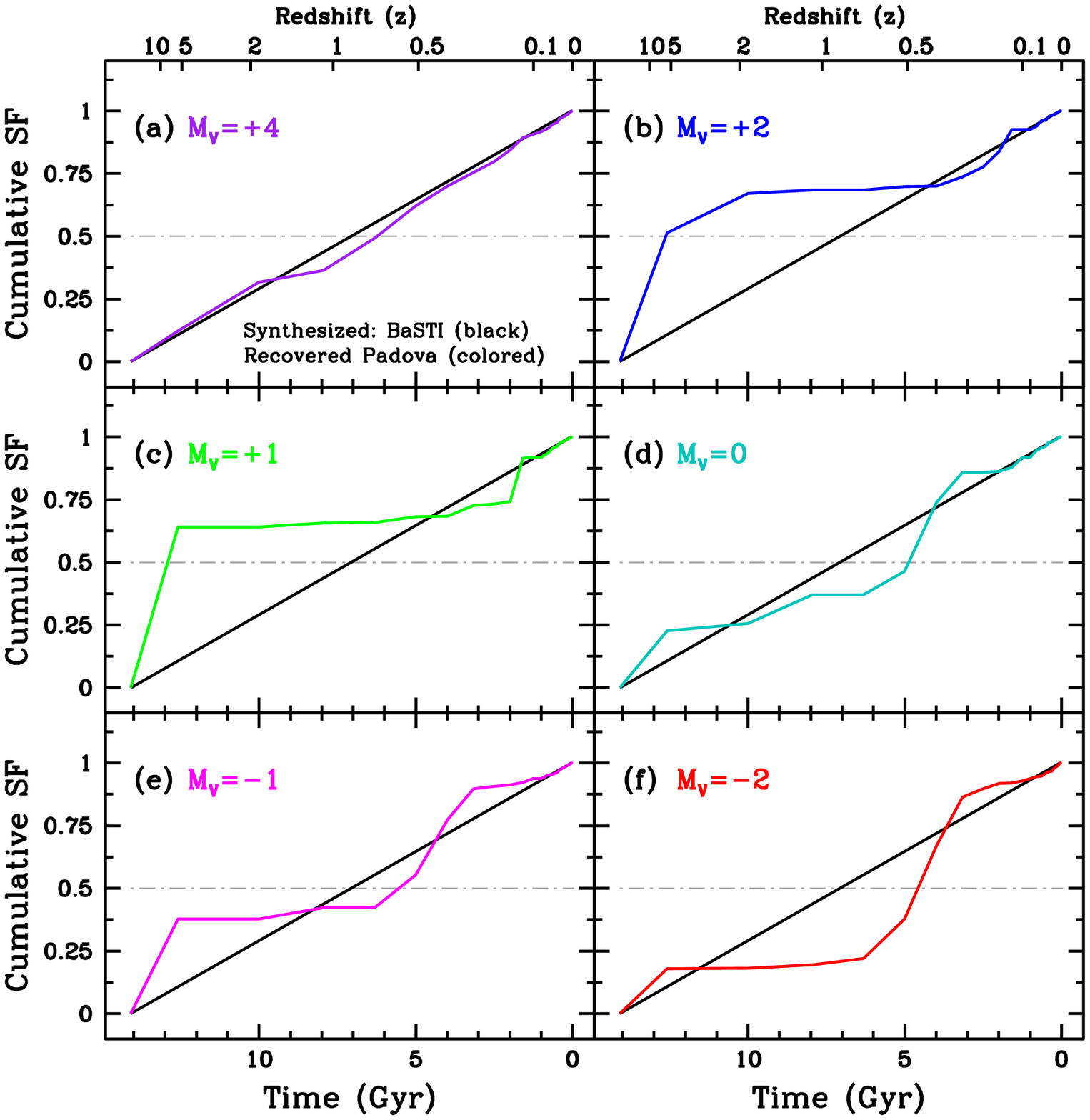}
\caption{\small{A test of the effects of photometric depth and stellar model differences on the accuracy of SFH recovery.  The black solid line is the input constant SFH, while the colored lines are the recovered SFHs at six select photometric depths.  Synthetic CMDs were generated using the BaSTI stellar models, a constant SFH, and a fixed metallicity. The SFHs of the synthetic CMDs were then recovered using identical parameters, and the Padova stellar evolution models.  The input and recovered SFHs show excellent agreement for the deepest CMD.  Discrepancies in the SFHs at shallower depths are indicative of the systematic effects introduced by the choice of stellar model.}}
\label{padbas_sf}
\end{center}
\end{figure}

In Figure \ref{padbas_sf}, we see the differences in the input (black lines) and recovered (colored lines) mean cumulative SFHs are more substantial than when identical models were used.  For the deepest CMD considered ($M_{V}=$$+$4),  the agreement between the input and recovered CMD is excellent.  Here, the ancient MS turnoff provides a reliable constraint on the ancient SFH, meaning the systematics in progressively younger time bins are also known more precisely.  For shallower CMDs that do not include the ancient MS turn-off, we see significant deviations between the recovered and input SFHs. These difference arise primarily in the treatment of such features of the horizontal branch, red clump, and RGB (see \citealt{gza05} for a more detailed discussion).

\begin{figure}[t]
\begin{center}
\epsscale{0.5}
\plotone{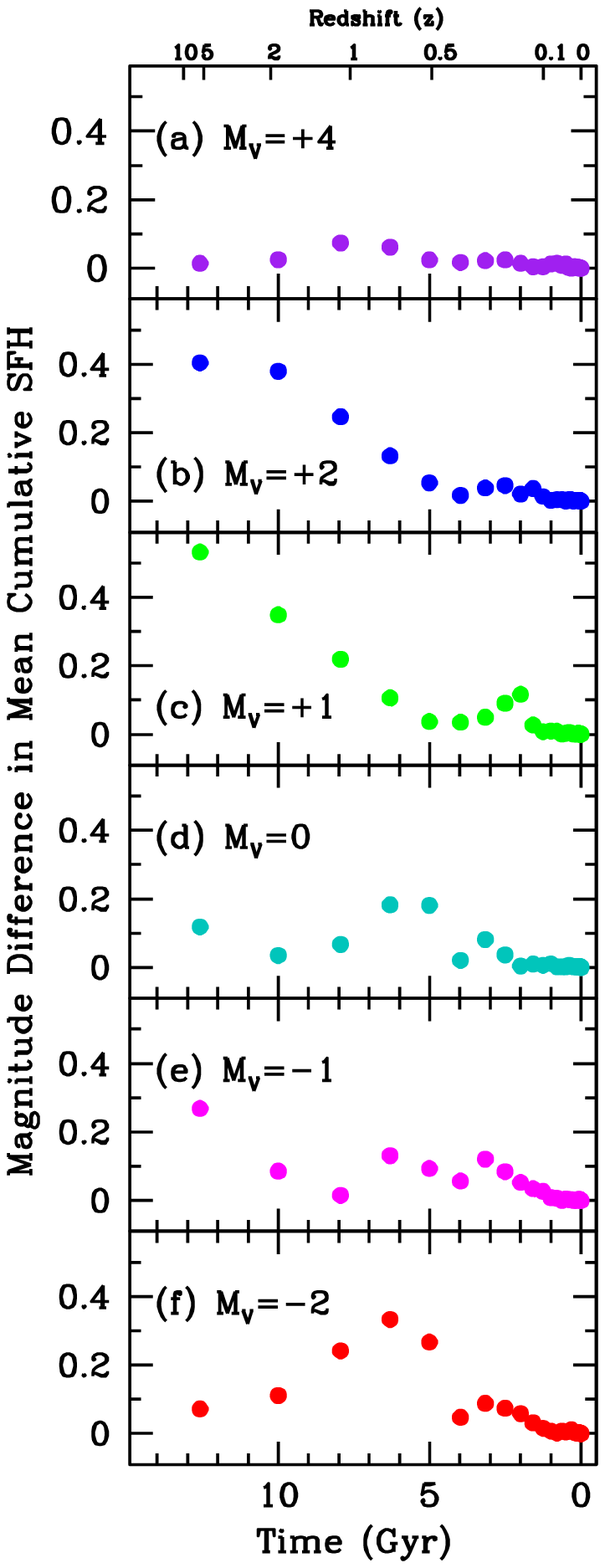}
\caption{\scriptsize{An estimate of the amplitude of systematic uncertainties on SFH between the BaSTI and Padova stellar evolution models.   Specifically, this figure highlights the absolute difference in the simulated input constant SFH (constructed with BaSTI) and the recovered SFH (solved with Padova) at six select photometric depths.  In general, deep CMD including the ancient MS turn-off ($M_{V}=$$+$4) show very little systematic difference.  SFHs from slightly shallower CMDs show SF at ancient times is generally overestimated by the Padova models, while the SFHs recovered from the shallowest CMDs place too much SF at intermediate epochs.  We caution that these results demonstrate the differences only between these two stellar models, and may not quantitatively reflect the differences between a given model and `truth', i.e., observed CMDs.}}
\label{diff_sf}
\end{center}
\end{figure}

We quantify the magnitude of the discrepancies between the Padova and BaSTI models by examining the absolute value of the differences in input and recovered cumulative SFHs (Figure \ref{diff_sf}). Of interest are the general order of magnitude variations, and not a specific point by point analysis.  In the case of the shallowest CMDs ($M_{V}=$$-$2, $-$1, 0), the ancient SF is preferentially recovered at intermediate ages, with a typical magnitude in the difference of input and recovered SFH of $\sim$ 20\%.  For the deeper CMDs ($M_{V}=$$+$1, $+$2), SF at the oldest times tends to be over-estimated by up to $\sim$ 40\%.  At all photometric depths, the systematic effects within the last $\sim$ 2 Gyr are $\lesssim$ 10\%. 

The exercises we have conducted demonstrate the importance of systematic uncertainties in SFH recovery.  We caution that our results are actually measuring the systematic differences between the Padova and BaSTI models, and only serve a a proxy for the difference between a given model and `truth', i.e., observed CMDs. That being said,  the methodology of this type of analysis provides a framework for exploring the effects of systematic uncertainties.  The primary limitation to this type of analysis is the number of models available that span the entire age/metallicity range needed to derive SFHs in nearby galaxies.  As the number of models that sample a broader parameter space increases, it will be possible to gain more leverage on the effects of systematic uncertainties. 

\newpage

\clearpage

\end{document}